\documentclass[12pt]{article} 
\def\cdate{{December 7, 2020}}
\def\myhead{{\it\small Ivan G. Avramidi: \mytitle}}
\def\mytitle{{Heat Kernel on Weyl Algebra}}
\usepackage{amsfonts}
\usepackage{amsmath}
\usepackage{latexsym}
\usepackage{amssymb}
\usepackage{mathptmx}
\usepackage[american]{babel}
\makeatletter
\def\timenow{%
\@tempcnta=\time \divide\@tempcnta by 60 \number\@tempcnta:\multiply
\@tempcnta by 60 \@tempcntb=\time \advance\@tempcntb by -\@tempcnta
\ifnum\@tempcntb <10 0\number\@tempcntb\else\number\@tempcntb\fi}
\newcounter{outputpage}
\renewcommand{\@oddhead}
{\stepcounter{outputpage}\myhead\hfill\hfill\theoutputpage}
\renewcommand{\@evenhead}
{\stepcounter{outputpage}\myhead\hfill\theoutputpage}
\renewcommand{\@oddfoot}
{\vbox{
\hrule
\vspace{3pt}
\hfil
{\scriptsize\textit{
\hfill\hfill\jobname.tex; \today; \timenow; p. \theoutputpage}}
\hfil
}}
\renewcommand{\@evenfoot}
{\vbox{
\hrule
\vspace{3pt}
\hfil
{\scriptsize\textit{
\hfill\hfill\jobname.tex; \today; \timenow; p. \theoutputpage
}}
\hfil
}}
\makeatother

\def\RR{{\mathbb R}} 
\def\CC{{\mathbb C}}

\def\gfrak{\mathfrak{g}} 
\def\hfrak{\mathfrak{h}}
\def\cA{{\cal A}}
 
\def\cG{\mathcal{G}}
\def\cD{{\cal D}}
\def\cH{{\cal H}}

\def\cL{{\cal L}}

\def\cF{{\cal F}}

\def\cR{\mathcal{ R}} 

\def\ad{\mathrm{ ad }} 
 
\def\tr{\mathrm{ tr\,}} 
\def\Tr{\mathrm{ Tr\,}}

\def\Real{\mathrm{Re\,}}

\def\be{\begin{equation}} 
\def\ee{\end{equation}} 
\def\bea{\begin{eqnarray}} 
\def\eea{\end{eqnarray}} 
\def\bed{\begin{definition}{\ }}
\def\eed{\end{definition}}
\def\bd{\begin{description}}
\def\ed{\end{description}}
\def\bc{\begin{center}}
\def\ec{\end{center}}
\newtheorem{theorem}{Theorem}
\newtheorem{lemma}{Lemma}
\newtheorem{corollary}{Corollary}

\newtheorem{definition}{Definition}
\def\sideremark#1{\ifvmode\leavevmode\fi\vadjust{\vbox to0pt{\vss
\hbox to 0pt{\hskip\hsize\hskip1em
\vbox{\hsize2cm\tiny\raggedright\pretolerance10000
\noindent #1\hfill}\hss}\vbox to8pt{\vfil}\vss}}}

\begin{document}

\begin{titlepage}

\thispagestyle{empty}
\null
\vspace{-10mm}
\hspace*{50truemm}{\hrulefill}\par\vskip-4truemm\par
\hspace*{50truemm}{\hrulefill}\par\vskip5mm\par
\hspace*{50truemm}{{\large\sc New Mexico Tech {\rm 
(\cdate)}}}\vskip4mm\par
\hspace*{50truemm}{\hrulefill}\par\vskip-4truemm\par
\hspace*{50truemm}{\hrulefill}
\par
\bigskip
\bigskip
\par
\par
\vspace{3cm}
\centerline{\huge\bf Heat Semigroups on Weyl Algebra}
\bigskip
\bigskip
\centerline{\Large\bf Ivan G. Avramidi
}
\bigskip
\centerline{\it Department of Mathematics}
\centerline{\it New Mexico Institute of Mining and Technology}
\centerline{\it Socorro, NM 87801, USA}
\centerline{\it E-mail: ivan.avramidi@nmt.edu}
\bigskip
\medskip
\vfill
{\narrower
\par


We study the algebra of semigroups of Laplacians
on the Weyl algebra. We consider first-order partial differential
operators $\nabla^\pm_i$
forming the Lie algebra $[\nabla^\pm_j,\nabla^\pm_k]=
i\cR^\pm_{jk}$ and $[\nabla^+_j,\nabla^-_k]
=i\frac{1}{2}(\cR^+_{jk}+\cR^-_{jk})$
with some anti-symmetric matrices $\cR^\pm_{ij}$
and define the corresponding Laplacians
$\Delta_\pm=g_\pm^{ij}\nabla^\pm_i\nabla^\pm_j$
with some positive matrices $g_\pm^{ij}$.
We show that the heat semigroups
$\exp(t\Delta_\pm)$ can be represented as a 
Gaussian average of the operators
$\exp\left<\xi,\nabla^\pm\right>$
and use these representations to compute
the product of the semigroups, $\exp(t\Delta_+)\exp(s\Delta_-)$
and the corresponding heat kernel.


\par}

\vfill
\hrule
\vspace{3pt}
\noindent
{\scriptsize\textit{
\hfill\hfill
\jobname.tex; \today; \timenow
}}

\end{titlepage}

\section{Introduction}
\setcounter{equation}0

Elliptic partial differential 
operators on manifolds play a crucial role in global analysis, spectral 
geometry and mathematical physics 
\cite{gilkey95,berger03,avramidi91,avramidi95,avramidi15}. 
The spectrum of elliptic 
operators does, of course, depend on the geometry of the manifold. Therefore, 
one can ask the question: ``Does the spectrum of an 
elliptic operator describe the geometry?''
It is well known now that the answer to this question is 
negative, that is, there are non-isometric manifolds that have the same 
spectrum (see, e.g. \cite{gordon92,schueth99,berger03}). 
Another area where elliptic operators are of great importantce is
the quantum field theory and quantum gravity
(see, for example, \cite{dewitt2003,grib94,wald94,parker09}).
In this setting the resolvent and the heat kernel of an elliptic
operator
enable one
to study the Green functions of quantum fields and the 
corresponding effective action. This is applied to study,
in particular, the creation of particles in strong time-dependent
gravitational and electromagnetic fields \cite{avramidi20b}.
An important tool to study the spectra of elliptic operators is the
heat semigroup and the associated heat kernel and its trace.
However, the heat trace describes only the eigenvalues of the operators
but not their eigensections.
That is why, it makes sense to study more general invariants of
partial differential operators, or, even, a collection of operators,
that might contain more information 
about the geometry of the manifold. Such invariants are not necessarily
spectral invariants that only depend on the eigenvalues of the operators;
they depend, rather, on both the eigenvalues and the eigensections.

In our paper \cite{avramidi16} we (with B. J. Buckman)
initiated the study of a new invariant 
of second-order elliptic partial
differential operators that we called {\it heat determinant}.
In our paper \cite{avramidi17} we studied so called
{\it relativistic heat trace} of a Laplace type operator $L$,
\be
\Theta_r(\beta)=\Tr\exp(-\beta\omega),
\ee
where  $\omega=\sqrt{L}$ and $\beta$ is a positive parameter
(that plays the role of the temperature),
as well as the  
{\it quantum heat traces},
\be
\Theta_{b,f}(\beta,\mu) 
=\Tr E_{b,f}\left[\beta\left(\omega-\mu\right)\right].
\ee
where $\mu$ is another real parameter (not necessarily positive,
that plays the role of the chemical potential)
and $E_{b,f}$ are functions defined by
\be
E_{b,f}(x)=\frac{1}{e^x\mp 1}.
\ee
These functions come from the quantum statistical physics
with the indices $b$ and $f$ standing for the 
bosonic and fermionic cases.
It was shown that such traces can be reduced by some
integral transform to the usual heat traces.

In the paper \cite{avramidi20a} 
we studied so called
{\it relative spectral invariants} of two operators $L_\pm$,
of the form
\be
\Psi(t,s) =
\Tr\left\{\exp(-tL_+)-\exp(-tL_-)\right\}
\left\{\exp(-sL_+)-\exp(-sL_-)\right\}.
\label{220ssb}
\ee
In our recent paper \cite{avramidi20b} we introduced another
invariant, called the {\it Bogolyubov invariant}, 
\bea
B_b(\beta)&=&\Tr
\Bigl\{E_f(\beta\omega_+)-E_f(\beta\omega_-)\Bigr\}
\Bigl\{E_b(\beta\omega_+)-E_b(\beta\omega_-)\Bigr\},
\label{311xxa}
\eea
and applied 
it to the study of particle creation in quantum field theory
and quantum gravity.
It was shown that these invariants can be reduced to the study
of the combined traces of the form
\be
X(t,s)=\Tr\left\{\exp(-tL_+)\exp(-tL_-)\right\}.
\label{16igax}
\ee
{\it The long term goal of this
project is to develop a comprehensive methodology for such invariants in the
same way as the theory of the standard heat trace invariants}.
The primary motivation for this study 
is spectral geometry and quantum field theory.


It is impossible to compute the combined traces (\ref{16igax}) exactly
for the general Laplace type operators $L_\pm$ on manifolds.
One can make some progress towards calculation of such traces
for operators and manifolds with some symmetries. This has been
initiated for scalar Laplacians with constant magnetic fields 
in our paper 
\cite{avramidi93,avramidi95} in $\RR^n$ and 
for scalar Laplacians on symmetric spaces. 
Finally, these ideas enabled us
to compute the heat trace for Laplacians on homogeneous bundles
over symmetric spaces in \cite{avramidi09}. 

In the present paper we study a very the product of the semigroups
of two operators $L_+$ and $L_-$,
\be
U(t,s)=\exp(-tL_+)\exp(-sL_-),
\ee
and the corresponding  kernels
by {\it using purely algebraic tools}.

The main idea of this approach can be described as follows.
Suppose that two Laplace type operators $L_\pm$ can be 
represented in the form
\be
L_\pm=G_\pm^{AB}\cL^\pm_A\cL^\pm_B,
\ee
where $G^{AB}$ is a positive symmetric matrix
and $\cL_A$ are some first-order partial differential
operators. Suppose further that the operators
$\cL_A^+,\cL_B^-$ form a closed Lie algebra,
say,
\bea
[\cL^+_A,\cL^+_B] &=& C_+^C{}_{AB}\cL^+_C+\cF^+_{AB},
\\{}
[\cL^-_A,\cL^-_B] &=& C_-^C{}_{AB}\cL^-_C+\cF^-_{AB},
\\{}
[\cL^+_A,\cL^-_B] &=& E_+^C{}_{AB}\cL^+_C
+E_-^C{}_{AB}\cL^-_C+\cR_{AB},
\eea
where $C_\pm^C{}_{AB}$, $\cF^\pm_{AB}$,
$E_\pm^C{}_{AB}$ and $\cR_{AB}$ are some constants
that we collectively call {\it curvatures} (since they
determine the commutators).
Here we assume that there are $N$ operators $\cL^+_A$
and $N$ operators $\cL^-_A$, so the indices
range over $1$ to $N$.
Then, by using purely algebraic methods one can try
to represent the heat semigroups in the form
\be
\exp(-tL_\pm)=\int_{\RR^N}d\xi\; \Phi_\pm(t,\xi)
\exp\left<\xi,\cL_\pm\right>,
\ee 
where $\left<\xi,\cL^\pm\right>=\xi^A\cL^\pm_A$
and $\Phi^\pm(t,\xi)$ are some functions
that depend on the curvatures.
Such representations were found in our papers
\cite{avramidi93,avramidi95,avramidi09}
in some special cases.

Then one can use this representation to
compute the convolution
\be
U(t,s)=\int_{\RR^{2N}}d\eta\,d\xi\;\Phi_+(t,\xi)\Phi_-(s,\eta)
\exp\left<\eta,\cL_+\right>\exp\left<\xi,\cL_-\right>.
\ee
And, finally, one can use the Campbell-Hausdorff type formulas
to compute the convolution of the operators
\be
\exp\left<\eta,\cL_+\right>\exp\left<\xi,\cL_-\right>
=\exp\left\{\left<F_+(\eta,\xi),\cL_+\right>
+\left<F_-(\eta,\xi),\cL_-\right>
\right\},
\ee
where $F_\pm(\eta,\xi)$ are some functions.
The main point of this idea is that {\it it is much easier
to compute the convolution of the exponentials of the first-order 
differential operators
that form some Lie algebra
than to compute the convolution of the exponential of the second-order 
differential operators}. 


For this idea to work we consider in the present paper a rather simple
very special {\it non-geometric setup} of operators on $\RR^n$,
such that they form a nilpotent algebra
described below. This problem has a direct application
to the creation of particles in time dependent
magnetic fields \cite{avramidi20b}.

In Sec.2 we describe the standard theory of Gaussian integrals 
in $\RR^n$ in the form that will be convenient for us later.
In particular, we introduced a Gaussian average of functions
on $\RR^n$ and study its properties. 
In Sec. 3 we study so called {\it Gaussian kernels}.
We show that the set of Gaussian kernel is a semigroup
with respect to the convolution and study some 
of its sub-semigroups.
In Sec. 4 we explore some of the well known formulas
related to the Campbell-Hausdorff series and prove
a couple of useful lemmas.
In Sec. 5 we introduce a real antisymmetric matrix
$\cR_{ij}$
that we call {\it curvature} study some canonical
functions of this matrix.

In Sec. 6 we consider the Heisenberg algebra and
its universal enveloping algebra.
We introduce a particular representation of 
the Heisenberg algebra 
related to the Weyl algebra
of differential operators with polynomial
coefficients,
$(\nabla_1,\dots,\nabla_n,
x^1,\dots,x^n, i)$, where 
\be
\nabla_j=\partial_j-\frac{1}{2}i\cR_{jk}x^k,
\label{nabla}
\ee
Given a positive matrix $g$ we introduce an operator
called the {\it Laplacian} by 
\be
\Delta_g=g^{ij}\nabla_i\nabla_j,
\label{delta}
\ee
where $g^{ij}$ is the inverse matrix.
{\it We consider all these operators acting in the space $C_0^\infty(\RR^n)$
of smooth functions with compact support}. Since
this space is dense in the Hilbert space $L^2(\RR^n)$,
the extension of these operators to the 
whole Hilbert space $L^2(\RR^n)$ is well defined. Moreover, the operators $\nabla_i$ are
{\it anti-self-adjoint} so that the Laplacian is self-adjoint.

In Sec. 7 we compute integrals of functions of the operators
$\nabla$. We prove the following theorem.

\begin{theorem}
\label{theorem2viaz}
Let $\cR=(\cR_{ij})$ be an anti-symmetric matrix and
$g=(g_{ij})$ be a positive symmetric matrix.
Let 
$D(t)=(D_{ij})$
be a symmetric matrix defined by
\be
D(t) 
=i\cR\coth\left(tg^{-1}i\cR\right)\,,
\ee
$T(t)$ be a matrix defined by 
\be
T(t)=D(t)+i\cR,
\ee
and $\Omega(t)$ be a function defined by
\bea
\Omega(t) &=& 
\det T(t)^{1/2}
=\det \left(g^{-1}\frac{\sinh(tg^{-1}i\cR)}{g^{-1}i\cR}\right)^{-1/2}.
\eea
Let $\nabla_i: C_0^\infty(\RR^n)\to C_0^\infty(\RR^n)$
be anti-self-adjoint first-order
partial differential operators 
of the form $\nabla_k=\partial_k-\frac{1}{2}i\cR_{kj}x^j$
acting on the space of smooth functions in $\RR^n$
forming the Lie algebra
\be
[\nabla_j,\nabla_k] = i\cR_{jk},
\ee
and $\Delta_g$ be the operator defined by
\be
\Delta_g=g^{ij}\nabla_i\nabla_j.
\ee
Then
\be
\exp(t\Delta_g) = (4\pi)^{-n/2}\Omega(t) 
\int\limits_{\RR^n} d\xi
\exp\left\{-\frac{1}{4}\left<\xi,D(t)\xi\right>
\right\}
\exp\left<\xi,\nabla\right>\,.
\label{858viax}
\ee
\end{theorem}

In Sec. 8 we consider two sets of such
{\it anti-self-adjoint} first-order
partial differential operators $\nabla_i^+$ and $\nabla_j^-$
and prove the following theorem.

\begin{theorem}
\label{theorem2in}
Let $\cR^\pm=(\cR^\pm_{ij})$ be two anti-symmetric matrices,
$g^\pm=(g^\pm_{ij})$ be two positive symmetric matrices and
let $D_\pm(t) $ be two
symmetric matrices defined  by
\be
D_\pm(t) = i\cR_\pm\coth(tg_\pm^{-1}i\cR_\pm).
\label{725igazx}
\ee
Let $\tilde\cF=(\tilde\cF_{AB})$ be a $2n\times 2n$ 
anti-symmetric matrix 
defined by
\bea
\tilde\cF
&=&\left(
\begin{array}{cc}
\cR^+ & \cR\\
\cR & \cR^-
\end{array}
\right),
\eea
where
\be
\cR=\frac{1}{2}\left(\cR^++\cR^-\right),
\label{84igaxm}
\ee
and $\tilde Q(t,s)=(\tilde Q_{AB})$
and
$\tilde\cG^{-1}(t,s)=(\tilde\cG^{AB})$
be the 
$2n\times 2n$ symmetric matrices
defined by
\bea
\tilde Q(t,s)
&=&\left(
\begin{array}{cc}
D_+(t) & -i\cR\\
i\cR & D_-(s)
\end{array}
\right),
\\
\tilde\cG^{-1}(t,s)
&=& \tanh^{-1}\left(\tilde Q^{-1}(t,s)i\tilde\cF\right)\left(i\cF\right)^{-1}.
\label{130igam}
\eea
Let $\nabla_i^\pm: C^\infty_0(\RR^n)\to C^\infty_0(\RR^n)$
be
anti-self-adjoint first-order
partial differential operators 
of the form
$\nabla^\pm_k=\partial_k-\frac{1}{2}i\cR^\pm_{kj}x^j$
acting on the space of smooth functions over $\RR^n$
forming the Lie algebra
\bea
[\nabla^+_i,\nabla^+_j] &=& i\cR^+_{ij},
\\{}
[\nabla^-_i,\nabla^-_j] &=& i\cR^-_{ij},
\\{}
[\nabla^+_i,\nabla^-_j] &=& i\cR_{ij},
\label{843viazy}
\eea
and $\Delta_\pm$ be two operators defined by
\be
\Delta_\pm 
=g_\pm^{ij}\nabla^\pm_i\nabla^\pm_j.
\ee
Let
 $(\tilde\cD_1,\dots,\tilde\cD_n,
\tilde\cD_{n+1},\dots,\tilde\cD_{2n})
=(\nabla^+_1, \dots, \nabla^+_{n},
\nabla^-_{1},\dots,\nabla^-_{n})$
and $\cH(t,s)$ be the operator defined by
\bea
\cH(t,s)=\left<\tilde\cD,\tilde\cG^{-1}(t,s)\tilde\cD\right>.
\label{831igaxy}
\eea
Then
\be
\exp(t\Delta_+)\exp(s\Delta_-)
=\exp\cH(t,s).
\ee
\end{theorem}

We should remark here that eq. (\ref{130igam}) should be understood
in terms of a power series; it is well defined even if the matrix
$\cF$ is not invertible.

We also prove the following theorem.
Let the matrices $\cR_\pm$, $g_\pm$, $D_\pm(t)$, 
and the operators $\nabla^\pm$ be defined as in
Theorem \ref{theorem2in}.
\begin{theorem}
\label{theorem3in}
Let $T_\pm(t)$, $D(t,s)$ and $Z(t,s)$ be the  
matrices defined by
\bea
T_\pm(t)&=&D_\pm(t)+i\cR,
\\
D(t,s)&=& D_+(t)+D_-(s),
\label{762igazx}
\\
Z(t,s) &=& D_+(t)-D_-(s)-2i\cR_-,
\label{491viacx}
\eea
and $\Omega(t,s)$ be a function defined by
\be
\Omega(t,s) = \det T_+(t)^{1/2}
\det T_-(s)^{1/2}
\det D^{-1/2}(t,s).
\label{2128iganx}
\ee
Let $H(t,s)$ be a symmetric matrix defined by
\be
H(t,s) = \frac{1}{4}\left(D(t,s)-Z^T(t,s)D^{-1}(t,s)Z(t,s)\right).
\label{859igaxy}
\ee
Let $\nabla_i$ and $X_j$ be the operators defined by
\bea
\nabla_i&=&\frac{1}{2}(\nabla^+_i+\nabla^-_i),
\label{351viax}
\\
X_i&=&\nabla^+_i-\nabla^-_i.
\label{352viay}
\eea
Then
\bea
&&\exp(t\Delta_+)\exp(s\Delta_-)
=
(4\pi)^{-n/2}
\Omega(t,s)
\exp\left<X,D^{-1}(t,s)X\right>
\label{490viaxc}
\\
&&\qquad\times
\int\limits_{\RR^{n}} d\alpha\,\exp\left\{
-\frac{1}{4}\left<\alpha,H(t,s)\alpha\right>
-\frac{1}{2}\left<\alpha,Z^T(t,s)D^{-1}(t,s)X\right>
\right\}
\exp\left<\alpha, \nabla\right>.
\nonumber
\eea
\end{theorem}

In Sec. 9 we compute the convolution of the heat kernels
and prove the following theorem.
Let the matrices $\cR_\pm$, $g_\pm$, $D_\pm(t)$, $T_\pm(t)$
and the operators $\nabla^\pm$ be defined as in
Theorem \ref{theorem3in}.

\begin{theorem}
\label{theorem4in}
Let $A_\pm(t,s)$ and $B(t,s)$ be matrices defined by
\bea
A_+(t,s) &=& D_+(t)-T_+^T(t)D^{-1}(t,s)T_+(t),
\\
A_-(t,s) &=& D_-(s)-T_-^T(s)D^{-1}(t,s)T_-(s),
\\
B(t,s)&=& T_+(t)D^{-1}(t,s)T_-(s),
\eea
and $S$ be a function defined by
\bea
S(t,s;x,x')&=&
\frac{1}{4}\left<x, A_+(t,s)x\right>
+\frac{1}{4}\left<x', A_-(t,s) x'\right>
-\frac{1}{2}\left<x,B(t,s)x'\right>.
\label{316iganx}
\eea
Then the kernel of the product of the semigroups is
\bea
U(t,s;x,x') &=&\exp(t\Delta_+)\exp(s\Delta_-)\delta(x-x'),
\nonumber\\
&=&\det\left(-\frac{S_{xx'}(t,s)}{2\pi}\right)^{1/2}
\exp\left\{-S(t,s;x,x')\right\}.
\label{820iganm}
\eea
\end{theorem}


\section{Gaussian Integrals}
\setcounter{equation}0

We will make extensive use of Gaussian
integrals.
We denote by $\left<\;,\;\right>$ the standard pairing in $\RR^n$.
Let $\gamma$ be a symmetric $n\times n$ matrix 
with positive definite real part. Then for any vector $A$
there holds (see, e.g. \cite{prudnikov83})
\be
\int_{\RR^n} d\xi\;\exp\left\{
-\frac{1}{4}\left<\xi,\gamma \xi\right>
+\left<A,\xi\right>\right\}
=(4\pi)^{n/2}\det \gamma^{-1/2}
\exp\left<A,\gamma^{-1}A\right>.
\label{gauss}
\ee
Let $S: \RR^n\to \CC$ be a quadratic polynomial
with positive definite real quadratic part.
Such a polynomial can always be written 
can be written in the form
\be
S(\xi)=S_0+\frac{1}{2}\left<S_\xi,S_{\xi\xi}^{-1}S_\xi\right>,
\ee
where $S_0$ is a constant,  $S_\xi$ and
$S_{\xi\xi}$ 
are the vector of first partial derivatives and the matrix 
of the second partial derivatives,
\bea
S_\xi &=& \left(\frac{\partial S}{\partial \xi^i}\right),
\\
S_{\xi\xi} &=& \left(\frac{\partial^2 S}{\partial\xi^i\partial\xi^j}\right).
\eea
Therefore, the Gaussian integral takes the form
\bea
\int_{\RR^n} d\xi\;\exp\left\{-S(\xi)\right\}
&=&
\det\left(\frac{S_{\xi\xi}}{2\pi}\right)^{-1/2}\exp(-S_0)
\nonumber\\
&=&
\det\left(\frac{S_{\xi\xi}}{2\pi}\right)^{-1/2}
\exp\left\{-S(\xi)+\frac{1}{2}\left<S_\xi,S_{\xi\xi}^{-1}S_\xi\right>
\right\}
\nonumber\\
&=&
\exp\left\{-\hat S(\xi)\right\},
\label{gauss1}
\eea
where
\be
\hat S=S-\frac{1}{2}\left<S_\xi,S_{\xi\xi}^{-1}S_\xi\right>
+\frac{1}{2}\tr\log\left(\frac{S_{\xi\xi}}{2\pi}\right);
\ee
which can be evaluated at an {\it arbitrary} point $\xi$.

Let $t>0$ be a positive parameter.
We introduce a one-parameter family of 
{\it Gaussian averages} of functions $f: \RR^n\to \CC$
by
\be
\left<f(\xi)\right>_t
=(4\pi t)^{-n/2}(\det\gamma)^{1/2}\int\limits_{\RR^n} d\xi
\exp\left\{-\frac{1}{4t}\left<\xi, \gamma\xi\right>\right\}
f(\xi).
\label{27igax}
\ee
Notice that this average depends on the parameter $t$
and
\be
\left<f(\xi)\right>_t=\left<f\left(\sqrt{t}\,\xi\right)\right>_1.
\ee
Therefore, for any smooth function, as $t\to 0^+$,
\be
\left<f(\xi)\right>_0=f(0).
\label{29igax}
\ee
By integration by parts we get useful equations
\bea
\left<\xi^if(\xi)\right>_t
&=&2t\gamma^{ij}\left<\frac{\partial}{\partial \xi^j}f(\xi)\right>_t,
\label{835viax}
\\
\left<\xi^j\xi^if(\xi)\right>_t
&=&
2t\gamma^{ij}\left<f(\xi)\right>_t
+4t^2\gamma^{jk}\gamma^{im}
\left<\frac{\partial}{\partial \xi^m}
\frac{\partial}{\partial \xi^k}f(\xi)\right>_t,
\label{835iga}
\eea
where $\gamma^{ij}$ is the inverse of the matrix $\gamma_{ij}$. 
Here and everywhere below, we denote the elements
of the inverse matrix by the same letter with upper indices.
Also, it is easy to see that
\bea
\partial_t \left<f(\xi)\right>_t
&=& \frac{1}{2t}\left<\xi^i\frac{\partial}{\partial \xi^i}f(\xi)\right>_t
\nonumber\\
&=&\left<\Delta_\xi f(\xi)\right>_{t},
\label{222cbxt}
\eea
where
\be
\Delta_\xi=\gamma^{ij}\frac{\partial}{\partial\xi^{i}}
\frac{\partial}{\partial\xi^j}.
\ee

We compute the Gaussian average of the function 
$\exp\left<\xi,ip\right>$. It satisfies the differential equation
\be
\partial_t\left<\exp\left<\xi,ip\right>\right>_t
=-\left<p,\gamma^{-1}p\right>\left<\exp\left<\xi,ip\right>\right>_t,
\ee
and, therefore,
\be
\left<\exp\left<\xi,ip\right>\right>_t
=\exp\left(-t\left<p,\gamma^{-1} p\right>\right).
\ee
By expanding both sides in the Taylor series we see
that Gaussian averages of odd order homogeneous polynomials vanish and
for even order homogeneous polynomials we obtain
\be
\left<\xi^{i_1}\dots\xi^{2k}\right>_t
=\frac{(2k)!}{k!}t^k\gamma^{(i_1i_2}\cdots\gamma^{i_{2k-1}i_{2k})}.
\label{229igan}
\ee
Therefore, for any analytic function the Gaussian average is
\be
\left<f(\xi)\right>_t
=\sum_{k=0}^\infty 
\frac{t^k}{k!}\Delta_\xi^k f(\xi)\Big|_{\xi=0}
=\exp\left(t\Delta_\xi\right)f(\xi)\Big|_{\xi=0}.
\ee
Notice that this average only contains the inverse matrix $\gamma^{-1}$.
So, strictly speaking, this average is defined also in the limiting 
case when the matrix $\gamma^{-1}$ is degenerate, that is,
has zero eigenvalues.

\section{Gaussian Kernels}
\setcounter{equation}0
 
Let $S$ be a quadratic polynomial on $\RR^n\times\RR^n$
of the form
\be
S(x,y) =
\frac{1}{4}\left<x,Ax\right>
-\frac{1}{2}\left<x,Cy\right>
+\frac{1}{4}\left<y,By\right>
-\frac{1}{2}\left<v,x\right>
-\frac{1}{2}\left<w,y\right>
+r,
\ee
where
$A$ and $B$ are real symmetric positive matrices,
$C$ is a complex matrix,
$v,w$ are some vectors and $r$ is a complex number.
We introduce the following notation for the derivatives
of the function $S$
\bea
S_1(x,y) &=& S_x(x,y)=\frac{1}{2}\left(Ax-Cy-v\right)
\\
S_2(x,y) &=& S_y(x,y)=\frac{1}{2}\left(-C^Tx+By-w\right)
\\
S_{11}(x,y)&=& S_{xx}(x,y)=\frac{1}{2}A,
\\
S_{12}(x,y)&=& S_{xy}(x,y)=-\frac{1}{2}C,
\\
S_{22}(x,y)&=& S_{yy}(x,y)=\frac{1}{2}B.
\eea

Let $\cA$ be the $2n\times 2n$ matrix
\be
\cA=\left(
\begin{array}{cc}
A & -C \\
-C^T & B\\
\end{array}
\right).
\ee
We assume that the matrix
$\Real\cA$
is non-negative, $\Real \cA\ge 0$.
This can be achieved by parametrizing the matrices $A$ and $B$
by
\bea
A &=& D^2,
\\
B &=& E^2,
\\
\Real C &=& D\Lambda E,
\eea
where $D$ and $E$ are real symmetric positive matrices
and $\Lambda$ is an orthogonal matrix,
that is,
\be
\Real \cA
=\left(
\begin{array}{cc}
D^2 & -D\Lambda E \\
-E\Lambda^T D & E^2\\
\end{array}
\right),
\ee
Then the quadratic part of the function $S$
is nonnegative;
indeed, in this case
\be
S(x,y)=
\frac{1}{4}||Dx-\Lambda Ey||^2
+\cdots.
\ee

Each such quadratic polynomial $S$ defines a function $U_S$ 
(that we call a {\it Gaussian kernel})
on $\RR^n\times \RR^n$ of the form
\be
U_S(x,y)=\det\left(-\frac{S_{12}}{2\pi}\right)^{1/2}
\exp\left\{-S(x,y)\right\}.
\ee
Let $\cG$ be the set of all Gaussian kernels.
We define the 
convolution of kernels by
\be
(U_S\circ U_{S'})(x,z)=\int_{\RR^n}dy U_S(x,y)U_{S'}(y,z).
\ee
It is easy to 
show that the convolution of Gaussian kernels is again
a Gaussian kernel, that is, the set $\cG$ is closed
under convolution. 
Since the convolution is associative, the set
$\cG$ is a semigroup. The group multiplication $*$
is defined by
\be
U_S\circ U_{S'}=U_{S * S'}.
\ee
It is not a group since it does not
have the identity and the inverses.
 
Notice that a Gaussian kernel $U_{\Sigma}$
with the function $\Sigma$ of the form
\be
\Sigma(x,y)=\frac{1}{4}\left<(x-y),g(x-y)\right>,
\ee
that is, $A=B=C=g$,
with $g$ a symmetric positive matrix
 and $v=w=r=0$,
plays the role of the asymptotic identity as
\be
U_{n\Sigma}(x,y)\stackrel{n\to\infty}{\longrightarrow}
\delta(x-y).
\ee
So, even though $\cG$ is not a group there is a sequence of 
kernels that converge to the identity. 

It is easy to see that the subset $\cG_0$ of Gaussian
kernels of the form $U_\Sigma$ is closed
under convolution and forms an Abelian
sub-semigroup
with the group multiplication defined
as follows: the function
$\tilde\Sigma=\Sigma*\Sigma'$
is defined by the matrix 
\be
\tilde g=\left(g^{-1}+g'^{-1}\right)^{-1}
=g'(g+g')^{-1}g.
\label{352igaw}
\ee

In general, the group multiplication has the 
following form.
Let $\tilde S=S* S'$; then
\bea
&&\tilde S(x,z)
= S(x,y)+S'(y,z)
\\
&&\qquad\qquad
-\frac{1}{2}\left<\left[S_2(x,y)+S'_1(y,z)\right],
\left(S_{22}+S'_{11}\right)^{-1}\left[S_2(x,y)+S'_1(y,z)\right]\right>,
\nonumber
\eea
with an {\it arbitrary} $y$, or, more explicitly,
\bea
\tilde S(x,z)
&=&
\frac{1}{4}\left<x,\left[A
-C(B+A')^{-1}C^T
\right]x\right>
+\frac{1}{4}\left<z,\left[B'
-C'^T(B+A')^{-1}C'
\right]
z\right>
\nonumber\\
&&
-\frac{1}{2}\left<x,C(B+A')^{-1}C'z\right>
-\frac{1}{2}\left<
\left[v
+C(B+A')^{-1}\left(w+v'\right)
\right],x\right>
\nonumber\\
&&
-\frac{1}{2}\left<
\left[w'
+C'^T(B+A')^{-1}\left(w+v'\right)
\right],z\right>
+r+r'
\nonumber\\
&&
-\frac{1}{4}\left<
\left(w+v'\right),
(B+A')^{-1}
\left(w+v'\right)
\right>.
\eea
Thus, finally, we the group transformation reads
\bea
\tilde A &=& A
-C(B+A')^{-1}C^T,
\\
\tilde B &=& B'
-C'^T(B+A')^{-1}C',
\\
\tilde C &=& C(B+A')^{-1}C',
\\
\tilde v &=&
v
+C(B+A')^{-1}\left(w+v'\right),
\\
\tilde w &=&
w'
+C'^T(B+A')^{-1}\left(w+v'\right),
\\
\tilde r &=&
r+r'
-\frac{1}{4}\left<
\left(w+v'\right),
(B+A')^{-1}
\left(w+v'\right)
\right>.
\eea

{\bf Remarks.}
Notice that this product is non-Abelian, in general.
The semigroup of Gaussian kernels $\cG$ has various subsemigroups;
these are the subsets closed under the group
multiplication.
\begin{enumerate}
\item
The subset $\cG_c$  with $C=0$.
In this case 
$\tilde A=A, \tilde B=B', \tilde v=v, \tilde w=w'$ and
\be
\tilde r =
r+r'
-\frac{1}{4}\left<
\left(w+v'\right),
(B+A')^{-1}
\left(w+v'\right)
\right>.
\ee

\item
The subset $\cG_2$ with
$v=w=r=0$, that is, with quadratic homogeneous
polynomial $S$. In this case
\bea
\tilde A &=& A
-C(B+A')^{-1}C^T,
\\
\tilde B &=& B'
-C'^T(B+A')^{-1}C',
\\
\tilde C &=& C(B+A')^{-1}C'.
\eea

\item
The subset $\cG_0$ with 
$C=B=A$ and $v=w=r=0$.
In this case the group transformation law
is Abelian; it
takes a particularly simple form,
\be
\tilde A^{-1} = A^{-1}+A'^{-1}.
\ee
This is exactly the transformation
(\ref{352igaw}) discussed above.

\end{enumerate}


\section{Campbell-Hausdorff Formula}
\setcounter{equation}0

We describe some of the well-known facts about the
Campbell-Hausdorff formula ({\it for a detailed exposition
see, e.g. \cite{bonfiglioli12}}).
Let $\gfrak$ be a Lie algebra.
For any operator $X\in \gfrak$ we define the operator
$\ad_X: \gfrak\to \gfrak$ by 
\be
\ad_X Y=[X,Y].
\ee
Let $P$ and $X$ be some operators in some Lie algebra $\gfrak$.
Our goal is to compute the product
$(\exp P)(\exp X)$.
We proceed rather formally.

We consider a 
smooth path,
$Q_t$,  $t\in[0,1]$ in a Lie algebra.
We prove a useful lemma.
 
\begin{lemma}
The derivative of the exponential $\exp Q_t$ is given by
\bea
\partial_t\exp Q_t &=& \exp(-Q_t)
\left\{\frac{1-\exp(-\ad_{Q_t})}{\ad_{Q_t}}\partial_t Q_t\right\},
\label{7135viaxc1}
\\
&=&
\left\{\frac{\exp(\ad_{Q_t})-1}{\ad_{Q_t}}\partial_t Q_t\right\}
\exp Q_t.
\label{41igaz}
\eea
\end{lemma}
{\it Proof.}
Let $F(t,s)$ be a function defined by
\be
F(t,s)=\exp(-sQ_t)\partial_t\exp(sQ_t),
\ee
so that 
$F(t,0)=0$.
Then it is not difficult to see that
\bea
\partial_s F 
&=&
\exp(-sQ_t)(\partial_t Q)\exp(sQ_t)
=\exp\left(-s\ad_{Q_t}\right)\partial_tQ.
\eea
Therefore, by integrating over $s$ from $0$ to $1$ 
we get
\be
F(t,1)=\int_0^1 ds \exp\left(-s\ad_{Q_t}\right)\partial_t Q
=\frac{1-\exp(-\ad_{Q_t})}{\ad_{Q_t}}\partial_t Q,
\ee
which proves (\ref{7135viaxc1}). Eq. (\ref{41igaz}) is proved similarly.
$\Box$

The expression in (\ref{7135viaxc1}) and (\ref{41igaz}) are understood as
power series, that is,
\be
\exp(-Q_t)\partial_t\exp Q_t=\sum_{k=0}^\infty \frac{(-1)^k}{(k+1)!}\ad^{k}_{Q}\partial_t Q_t
=\partial_tQ_t+\mbox{commutators},
\ee
so, in some sense, this is a ``non-commutative derivative''.
For a more rigorous treatment one would need convergence in some 
sufficiently strong topology ({\it for details
see, e.g. \cite{bonfiglioli12}}).


\begin{corollary}
The operators $Q_1$ and $Q_0$ are related by
\be
Q_1 = Q_0 
+\int_0^1 dt\;\frac{\ad_{Q_t}}{1-\exp(-\ad_{Q_t})}
A(t),
\label{46igaz}
\ee
where
\be
A(t)=\exp(-Q_t)\partial_t\exp Q_t.
\label{}
\ee
\end{corollary}
{\it Proof.}
This follows directly from the differential equation
(\ref{7135viaxc1}).
$\Box$

This enables one to prove the general Campbell-Haussdorff theorem.
Let $P$ and $X$ be two operators in some Lie algebra.

\begin{theorem}
Let $\psi(z)$ be a function defined by
\be
\psi(z)=\frac{z}{z-1}\log z.
\ee 
Then
\be
(\exp P)\exp X=\exp V,
\ee
where
\be
V = P 
+\int_0^1 dt\;\psi\Big(\exp(\ad_P)\exp(t\ad_X)\Big)X.
\ee
\end{theorem}
{\it Proof.}
Let $Q_t$ be a one-parameter family of operators defined by
\be
\exp Q_t=(\exp P)\exp(t X),
\ee
so that $Q_0=P$ and $Q_1=V$.
It is easy to see that in this case
\be
A(t)=\exp(-Q_t)\partial_t (\exp Q_t)=X
\ee
Therefore, by using (\ref{46igaz}) we get
\be
V = P 
+\int_0^1 dt\;\frac{\ad_{Q_t}}{1-\exp(-\ad_{Q_t})}X.
\ee
Next, we can rewrite this in the form
\be
\frac{\ad_{Q_t}}{1-\exp(-\ad_{Q_t})}
=\frac{\exp(\ad_{Q_t})}{\exp(\ad_{Q_t})-1}\ad_{Q_t}
=\psi\left(\exp(\ad_{Q_t})\right).
\ee
Further, by using the fact that 
\be
\exp(\ad_{Q_t}) = \exp(\ad_P)\exp(t\ad_X).
\ee
we obtain the result.
$\Box$

Of course, this theorem is understood in terms of a power
series in the operators $P$ and $X$ by using the Taylor series
\be
\psi(z)=\frac{z}{z-1}\log z
=1-\sum_{k=0}^\infty\frac{1}{k(k+1)}(1-z)^k.
\ee 
and by rescaling the operators $P\mapsto sP, X\mapsto sX$ and expanding
everything in powers of $s$.

\begin{lemma}
Suppose that for any positive integer $k$
the operator $X$
commutes with all commutators $[P,\dots,[P,X]\dots]$,
that is, 
\be
\ad_X\ad_P^k X=0.
\ee
Then
\bea
\exp(P+X)
&=&\exp\left\{\frac{1-\exp(-\ad_{P})}{\ad_{P}}X\right\}\exp P
\nonumber\\
&=&
(\exp P)\exp\left\{\frac{\exp(\ad_{P})-1}{\ad_{P}}X\right\}.
\eea
\end{lemma}
{\it Proof.}
Let $Q_t=P+tX$. Then $\partial_t Q_t=X$ and by using (\ref{7135viaxc1})
we get
\be
A(t)=\exp(-Q_t)\partial_t\exp Q_t
=\frac{1-\exp(-\ad_{Q_t})}{\ad_{Q_t}}X.
\ee
We obviously have
\be
\ad_{Q_t}=\ad_P+t\ad_X.
\ee
Therefore, under the assumptitons of the lemma
\bea
\ad^k_{Q_t}X &=& (\ad_P+t\ad_X)^k X=\ad_P^kX,
\eea
and, therefore, the operator $A(t)$ does not depend on $t$,
\be
A(t)=\exp(-Q_t)\partial_t\exp Q_t=
\frac{1-\exp(-\ad_{P})}{\ad_{P}}X.
\ee
Therefore, we can integrate this differential equation
with the initial condition
$\exp(Q_t)|_{t=0}=\exp P$
to obtain
\be
\exp Q_t
=\exp\left\{t\frac{1-\exp(-\ad_{P})}{\ad_{P}}X\right\}\exp P;
\ee
and by setting $t=1$ we prove the lemma.
The second equation is obtained by taking the inverse of the first
and changing the signs of $P$ and $X$.
$\Box$

\begin{corollary}
Suppose that for any positive integer $k$
the operator $X$
commutes with all commutators $[P,\dots,[P, X]\dots]$, 
that is, 
\be
\ad_X\ad^k_P X=0.
\ee
Then
\bea
\exp(X)\exp(P)
&=&\exp\left\{P+\frac{1-\exp(\ad_{P})}{\ad_{P}}X\right\}
\label{7146viac}\\
\exp(P)\exp(X)
&=&\exp\left\{P+\frac{\exp(-\ad_{P})-1}{\ad_{P}}X\right\}.
\eea
\end{corollary}
{\it Proof.} Let $Y=P+X$. 
This operator commutes with all commutators
$\ad^k_P Y$ and for any $k>1$
\be
\ad^k_P X=\ad^k_P Y,
\ee
and, therefore,
\be
\frac{1-\exp(-\ad_{P})}{\ad_{P}}X
=-P
+\frac{1-\exp(-\ad_{P})}{\ad_{P}}Y.
\ee
Therefore,
\be
\exp Y
=\exp\left\{-P
+\frac{1-\exp(-\ad_{P})}{\ad_{P}}Y\right\}\exp P.
\ee
Now, by multiplying by $\exp(-P)$ on the right and
changing the sign of $P$ we prove eq. (\ref{7146viac}). 
The second equation is obtained by taking the inverse.
$\Box$

In the case when the commutator $[A,B]$ commutes with both
operators $A$ and $B$, that is, $[B,[A,B]]=[A,[A,B]]=0$,
this reduces to the well known
special case of the Campbell-Hausdorff formula
\bea
(\exp A)(\exp B) &=& \exp\left(\frac{1}{2}[A,B]\right)\exp(A+B),
\label{285igam}
\eea
which also means 
\be
(\exp A)(\exp B) =\exp[A,B]
(\exp B)(\exp A).
\label{285igamx}
\ee

\section{Curvature}
\setcounter{equation}0

Let $M_n(\RR)$ be the algebra 
$n\times n$ matrices with the standard matrix product
and $\cR=(\cR_{ij})\in M_n(\RR)$ be a fixed real
antisymmetric matrix
that we call the {\it curvature}.
We define
a bilinear binary operation
(that we will call a $\cR$-bracket)
\be
\{\;,\;\}: M_n(\RR)\times M_n(\RR)\to M_n(\RR)
\ee
as follows: for any two
matrices $A=(A^{ij})$ and $B=(B^{ij})$
\be
\{A,B\}=A\cR B-B\cR A.
\label{52igam}
\ee
Obviously, this bracket is anti-symmetric,
$\{A,B\}=-\{B,A\}$
and satisfies the Jacobi identity
\be
\{A,\{B,C\}\}+\{B,\{C,A\}\}+\{C,\{A,B\}\}=0
\ee
Let $S_n\subseteq M_n(\RR)$ and $L_n\subseteq M_n(\RR)$ be the subspaces
of symmetric and anti-symmetric matrices. 
It is easy to see that they are closed under the
$\cR$-bracket which defines
the Lie brackets $\{\;,\;\}: S_n\times S_n\to S_n$
and $\{\;,\;\}: L_n\times L_n\to L_n$
turning them into Lie algebras of symmetric
and anti-symmetric matrices.

Let $\gamma=(\gamma_{ij})$ be a positive symmetric matrix. 
We will be considering analytic functions 
$f(\gamma^{-1}i\cR)$
of the matrix $\gamma^{-1}i\cR$.
It is easy to see that for any non-negative integer $k$ 
\be
\left[\left(\gamma^{-1}i\cR\right)^k\right]^T
=\gamma\left(-\gamma^{-1}i\cR\right)^k\gamma^{-1};
\ee
therefore, for any analytic function $f$ 
\be
\left[f(t\gamma^{-1}i\cR)\right]^T=\gamma f(-t\gamma^{-1}i\cR)\gamma^{-1};
\ee
hence, for any even function $f$
the matrices $\gamma f(\gamma^{-1}i\cR)$
and $f(\gamma^{-1}i\cR)\gamma^{-1}$
are symmetric.
More generally, let $\Lambda$ be a nondegenerate
matrix;
then for any analytic function we have
\be
f(t\gamma^{-1}i\cR)\gamma^{-1}=\Lambda f(t\gamma'^{-1}i\cR') \gamma'^{-1}\Lambda^T,
\ee
where
\bea
\gamma'&=& \Lambda^{T}\gamma\Lambda, 
\\
\cR'&=& \Lambda^T\cR\Lambda.
\eea
The matrix $\Lambda$ is arbitrary and can be
chosen from convenience.

We can always write the matrix $\gamma$ in the form
\be
\gamma^{-1}=\omega\omega^T,
\ee
where $\omega$ is some non-degenerate matrix.
Then the matrix $\gamma^{-1}i\cR$ has
the following canonical form
\be
\gamma^{-1}i\cR=\omega\sum_{\alpha=1}^{m} B_\alpha E_\alpha\omega^{-1},
\label{311igaz}
\ee
where $m\le n/2$, $B_\alpha$ are some real invariants,
and $E_\alpha$ are irreducible anti-symmetric matrices
satisfying
\bea
E_{\alpha}^2 &=& P_\alpha,
\\
E_\alpha E_\beta &=& 0, \qquad\mbox{for}\qquad \alpha\ne\beta,
\eea
with the corresponding symmetric projections $P_\alpha$
satisfying
\bea
P_{\alpha}^2 &=& P_\alpha,
\\
P_\alpha P_\beta &=& 0, \qquad\mbox{for}\qquad \alpha\ne\beta,
\\
\tr P_\alpha &=&2.
\eea

Therefore, for any even analytic function $f$ we have
\be
f(t\gamma^{-1}i\cR)=\omega
\left\{
f(0)I+\sum_{\alpha=1}^{m} [f(tB_\alpha)-f(0)]P_\alpha
\right\}\omega^{-1}.
\ee


In this paper we will be using extensively two even functions
\bea
\Phi(z) &=& \frac{\tanh^{-1}(z)}{z} 
=\frac{1}{2z}\log\left(\frac{1+z}{1-z}\right)
\nonumber\\
&=&\sum_{k=0}^\infty \frac{1}{2k+1}z^{2k}
=1+\frac{1}{3}z^2+\cdots,
\label{219iga}
\\
\Psi(z) &=& z\coth z 
\nonumber\\
&=& \sum_{k=0}^\infty \frac{2^{2k}B_{2k}}{(2k)!}z^{2k}
=1+\frac{1}{3}z^2+\cdots,
\label{220iga}
\eea
where $B_n$ are Bernoulli numbers.
Note that the function $\Phi(z)$ is analytic
with cuts along the real axis from $-\infty$ to $-1$ and
from
$1$ to $\infty$, whereas the function $\Psi(z)$ is 
meromorphic with simple poles on the imaginary axis
at $z=2\pi i n$, where $n$ is a non-zero integer, $n\ne 0$.
That is, the function $\Phi(x)$ is singular at $x=\pm 1$
and $\Psi(x)$ is well defined on the whole real line.
These functions are related by
\be
\Psi(z)=\Phi(\tanh z).
\label{222igaz}
\ee
In particular, we have
\bea
\Phi(t\gamma^{-1}i\cR)
=\omega\left\{
I+\sum_{\alpha=1}^{m}\left[\frac{\tanh^{-1}(tB_\alpha)}{tB_\alpha}-1\right]P_\alpha
\right\}\omega^{-1},
\\
\Psi(t\gamma^{-1}i\cR)
=\omega\left\{
I+\sum_{\alpha=1}^{m}[tB_\alpha\coth(tB_\alpha)-1]P_\alpha
\right\}\omega^{-1}
\,.
\eea

\section{Weyl Algebra}
\setcounter{equation}0

The Lie algebra $\hfrak_n$ of the Heisenberg group $H_{2n+1}$
is generated by the $(2n+1)$ operators
$(p_1,\dots,p_n, x^1,\dots,x^n,i)$
satisfying the commutation relations
\bea
[p_k, x^j] &=&i\delta^j_k,
\\{}
[p_k,p_j] &=& [x^k,x^j] = [p_k,i]=[x^k,i]=0.
\eea
We will just call it the {\it Heisenberg algebra}.
We will use another basis 
of the Heisenberg algebra 
$(P_1,\dots,P_n, x^1,\dots,x^n,i)$ 
defined by
\be
P_k=p_k+\frac{1}{2}\cR_{kj}x^j;
\ee
satisfying the commutation relations
\bea
[P_k,x^j]&=&i\delta^j_k,
\\{}
[P_k, P_j] &=& -i\cR_{kj}\,,
\label{45igaz}
\\{}
[x^k,x^j]&=&[P_k,i]=[x^k,i]=0.
\label{p-comm}
\eea
Obviously, the operators $(P_j,i)$ form a subalgebra
of the Heisenberg algebra.

Its universal enveloping algebra $U(\hfrak_n)$ is the set
of all polynomials in these operators subject to these
commutation relations.
Let $(x^1,\dots,x^n)$ be the coordinates of the Euclidean space $\RR^n$
and $(\partial_1,\dots,\partial_n)$ be the corresponding partial derivatives.
Obviously, the Heisenberg algebra can be represented by
the first order differential operators $p_k=i\partial_k$.
Then the universal enveloping algebra $U(\hfrak_n)$ is simply
the ring of all differential operators with polynomial coefficients,
also called the {\it Weyl algebra} $A_n$.
The operators $P_k$ then have the form $P_k=i\nabla_k$ with
\be
\nabla_k=\partial_k-\frac{1}{2}i\cR_{kj}x^j.
\label{67igax}
\ee

More precisely, we consider the partial derivative
operators $\partial_i$ acting on the space $C^\infty_0(\RR^n)$
of {\it smooth functions of compact support}. Recall that
this space is dense in the Hilbert space $L^2(\RR^n)$,
which defines the extension of these operators to the 
whole Hilbert space $L^2(\RR^n)$. 
Moreover, the derivatives operators $\partial_i$ are
unbounded (essentially)
{\it anti-self-adjoint} operators in this Hilbert space.
Then the operators $\nabla_i$ are {\it anti-self-adjoint} and
satisfy the commutation relations
\bea
[\nabla_k,x^j]&=&\delta^j_k,
\\{}
[\nabla_k,\nabla_j] &=& i\cR_{kj}\,,
\label{45igazx}
\\{}
[\nabla_k,i]&=&[x^k,i]=0.
\label{nabla-comm}
\eea

We consider analytic functions $f(i,x,\nabla)$
defined by power series with the coefficients 
in the Weyl algebra, $A_n$.
We will be mostly interested in analytic functions
functions $f(i,\nabla)$
that only depend on the 
operators $\nabla_j$ and $i$ but not on $x^k$.

Note that for the anti-self-adjoint operators $\nabla_i$,
the operator $\exp\left<\xi,\nabla\right>$ 
with an arbitrary $\xi\in\RR^n$
is well defined. Since the operators
$\left<\xi,i\cR x\right>$
and $\left<\xi,\partial\right>$ commute we have
the following lemma.
\begin{lemma}
\label{lemma2igaz}
Let $\nabla_i: C_0^\infty(\RR^n)\to C_0^\infty(\RR^n)$ 
be the first-order partial differential operators 
acting on the space of smooth functions
of the form
$\nabla_k=\partial_k-\frac{1}{2}i\cR_{kj}x^j$.
Then the operator $\exp\left<\xi,\nabla\right>: C^\infty_0(\RR^n)\to C^\infty_0(\RR^n)$
is an isometry.
\end{lemma}
{\it Proof.}
For any $\xi,x\in\RR^n$ there holds
\bea
\exp\left<\xi,\nabla\right>
&=&
\exp\left(-\frac{1}{2}\left<\xi,i\cR x\right>\right)
\exp\left<\xi,\partial\right>\,.
\label{612igaxm}
\eea
Therefore, it acts on a function $f\in C^\infty_0(\RR^n)$
by
\bea
\left(\exp\left<\xi,\nabla\right>f\right)(x)
&=&
\exp\left(-\frac{1}{2}\left<\xi,i\cR x\right>\right)
f(x+\xi)\,.
\label{612igaxb}
\eea
Obviously, the image of a smooth function of compact support
is again a smooth function of compact support;
also, this action preserves the $L^2$ norm.
$\Box$

Further, we will need the following lemma.

\begin{lemma}
\label{lemma1igaz}
For any $\eta,\xi\in\RR^n$ there holds
\bea
\exp\left<(\xi+\eta),\nabla\right>
&=&
\exp\left<\eta,\left(\nabla-\frac{1}{2}i\cR\xi\right)\right>
\exp\left<\xi,\nabla\right>\,.
\label{612igax}
\eea
\end{lemma}
{\it Proof.} 
By using the Campbell-Hausdorff formula (\ref{285igam})
we have
\be
\exp\left<(\xi+\eta),\nabla\right>
=
\exp\left\{
\left<\eta,\nabla\right>
-\frac{1}{2}[\left<\eta,\nabla\right>,\left<\xi,\nabla\right>]
\right\}
\exp\left<\xi,\nabla\right>.
\ee
Now, by using the commutator (\ref{45igaz})
we obtain the result.
$\Box$

\begin{corollary}
\label{lemma1viaz}
The partial derivatives of the operator $\exp\left<\xi,\nabla\right>$ are
\bea
&&\frac{\partial}{\partial \xi^{i_1}}\cdots
\frac{\partial}{\partial \xi^{i_k}}
\exp\left<\xi,\nabla\right>
\\
&&\qquad
=
\left(\nabla_{(i_1}-\frac{1}{2}i\cR_{(i_1|j_1|}\xi^{j_1}\right)\cdots
\left(\nabla_{i_k)}-\frac{1}{2}i\cR_{i_k)j_k}\xi^{j_k}\right)
\exp\left<\xi,\nabla\right>,
\nonumber
\eea
in particular,
\bea
\frac{\partial}{\partial \xi^i}\exp\left<\xi,\nabla\right>
&=&\left(\nabla_i-\frac{1}{2}i\cR_{ij}\xi^j\right)\exp\left<\xi,\nabla\right>
\label{829viaz}
\\
\frac{\partial}{\partial \xi^i}\frac{\partial}{\partial \xi^j}\exp\left<\xi,\nabla\right>
&=&\left(\nabla_{(j}-\frac{1}{2}i\cR_{(j|k|}\xi^k\right)
\left(\nabla_{i)}-\frac{1}{2}i\cR_{i)m}\xi^m\right)
\exp\left<\xi,\nabla\right>.
\label{29iga}
\eea
\end{corollary}
{\it Proof.} This is proved by expanding (\ref{612igax}) in Taylor series in $\eta$.

Alternatively, eq. (\ref{829viaz}) this can also be proved following 
\cite{avramidi93,avramidi15}.
Let
\be
F_i(t)=-\exp\left(t\left<\xi,\nabla\right>\right)
\frac{\partial}{\partial\xi^i}\exp\left(-t\left<\xi,\nabla\right>\right).
\ee
This operator satisfies the equation
\be
\partial_t F_i(t) = \ad_{\left<\xi,\nabla\right>}F_i(t)
+\nabla_i,
\ee
with the initial condition $F_i(0)=0$.
The solution of this equation is
\be
F_i(t)=\frac{\exp\left(t\ad_{\left<\xi,\nabla\right>}\right)-1}
{\ad_{\left<\xi,\nabla\right>}}\nabla_i
=\sum_{k=1}^\infty \frac{t^k}{k!}
\left(\ad_{\left<\xi,\nabla\right>}\right)^{k-1}\nabla_i,
\ee
The function $F_i(t)$ can be evaluated
by expanding it in the Taylor series.
First, we compute
\bea
\ad_{\left<\xi,\nabla\right>}\nabla_i&=&i\cR_{ij}\xi^j,
\\
\left(\ad_{\left<\xi,\nabla\right>}\right)^k\nabla_i&=&0,
\qquad\mbox{for any}\ k\ge 2;
\eea
therefore,
\be
F_i(t)=t\nabla_i-\frac{1}{2}t^2i\cR_{ij}\xi^j.
\ee

Next, we have
\bea
\frac{\partial}{\partial\xi^i}\exp\left(t\left<\xi,\nabla\right>\right)
&=&F_i(t)\exp\left(t\left<\xi,\nabla\right>\right)
\nonumber\\
&=&-\exp\left(t\left<\xi,\nabla\right>\right)F_i(-t).
\label{829viax}
\eea
By setting $t=1$ in (\ref{829viax}) we get eq. (\ref{829viaz}). 
Eq. (\ref{29iga}) follows by differentiation and
symmetrization.
$\Box$


Given a real symmetric positive matrix $g=(g_{ij})$ we define the operator
(that we call the {\it Laplacian})
\be
\Delta_g=\left<\nabla,g^{-1}\nabla\right>=g^{ij}\nabla_i\nabla_j,
\ee
where $g^{-1}=(g^{ij})$ is the inverse of the matrix $g$.
We can always write the matrix $g^{-1}$ in the form
$g^{-1}=\omega\omega^T$ so that
so that
\be
\Delta_g=\left<\nabla',\nabla'\right>,
\ee
where
\be
\nabla'=\omega^T\nabla.
\ee
These operators satisfy the same commutation relations
(\ref{45igaz})
with 
\bea
\cR'=\omega^T\cR\omega.
\eea
Furthermore, we can still transform the operators
$\nabla'$ by an orthogonal transformation 
\be
\tilde\nabla=O^T\nabla'=O^T\omega^T\nabla
\ee
with the orthogonal matrix $O$
so that
$\Delta_g=\left<\tilde\nabla,\tilde\nabla\right>$
and the corresponding curvature
\be
\tilde\cR=O^T\omega^T\cR\omega O,
\ee
to bring the matrix $\tilde\cR$ to a canonical form. 
Therefore, without loss of generality we can restrict 
in the following to the case $g=I$ and 
the matrix $\cR$ in the canonical form (\ref{311igaz});
so, we will drop the prime and the tilde below
and just assume that $g=I$ and $\cR$ 
has the form (\ref{311igaz}).


Our goal is computing the product of the semigroups
$
\exp(t\Delta_{g_+})\exp(s\Delta_{g_-})
$
for two symmetric positive matrices $g_\pm$. 

\begin{lemma}
\label{lemma0}
The set of Laplacian operators is closed under commutation.
That is, the commutator of Laplacians $\Delta_{g_+}$
and $\Delta_{g_-}$ is again a Laplacian
\be
[\Delta_{g_+},\Delta_{g_-}]
=2i\Delta_G,
\ee
where $G^{-1}={\{g_+^{-1},g_-^{-1}\}}=g_+^{-1}\cR g_-^{-1}
-g_-^{-1}\cR g_+^{-1}$ is given by the bracket
$\{\;,\;\}$ defined by (\ref{52igam}), that is,
$G^{ij}=g_+^{ik}\cR_{km}g_-^{mj}
-g_-^{ik}\cR_{km}g_+^{mj}$.
\end{lemma}
{\it Proof.}
We compute the commutator of these operators 
with the operators $\nabla_i$,
\be
[\nabla_i,\Delta_{g_+}]=2i\cR_{ij}g_+^{jk}\nabla_k.
\ee
The statement follows.
$\Box$

By using this lemma we have an immediate corollary.
\begin{corollary}
\label{lemma1}
The operators $\Delta_{g_+}$
and $\Delta_{g_-}$ commute if and only if
the matrices $g_+$ and $g_-$ satisfy the condition
\be
\{g_+^{-1}, g_-^{-1}\}=0.
\label{23viax}
\ee
\end{corollary}



\section{Noncommutative Gaussian Integrals}
\setcounter{equation}0

Let $f(\xi)$ be a function depending on the operators $\nabla_i$
and $\left<f(\xi)\right>_t$ be the Gaussian average defined by
(\ref{27igax}).
Suppose that we can find  operators $L(t)$ such that
this average satisfies the differential equation
\be
\partial_t  \left<f(\xi)\right>_t = L(t)  \left<f(\xi)\right>_t.
\ee
Recall that $\left<f(\xi)\right>_0=f(0)$, which
serves as the initial condition.
Then, if the operators $L(t)$ and $L(s)$ commute for any $t$ and $s$,
one can solve this equation to obtain
\be
\left<f(\xi)\right>_t
=\exp\left\{\int_0^t d\tau\; L(\tau)\right\}f(0).
\ee

We use this idea to study the Gaussian average
of the operator $\exp\left<\xi,\nabla\right>$,
\be
\left<\exp\left<\xi,\nabla\right>\right>_t
=(4\pi t)^{-n/2}(\det\gamma)^{1/2}\int\limits_{\RR^n} d\xi
\exp\left\{-\frac{1}{4t}\left<\xi, \gamma\xi\right>\right\}
\exp\left<\xi,\nabla\right>.
\label{237igax} 
\ee
This integral converges for {\it any bounded} operators $\nabla_i$.
It is not difficult to see that this integral converges if the
operators $\nabla_i$ are {\it anti-self-adjoint}, {\it even if they
are unbounded}. In this case the operator $\exp\left<\xi,\nabla\right>$
is an isometry and we just take a Gaussian average of isometries,
which is a bounded operator.
This can also be seen by using the spectral resoltuion of these
operators.
In the case when the operators $\nabla_i$ are given by
(\ref{67igax}) we can prove the lemma.

\begin{lemma}
The operator
$K=\left<\exp\left<\xi,\nabla\right>\right>_t: C_0^\infty(\RR^n)\to C_0^\infty(\RR^n)$
is well defined with the integral kernel
\be
K(t;x,x')=(4\pi t)^{-n/2}(\det\gamma)^{1/2}
\exp\left\{-\frac{1}{4t}\left<(x-x'), \gamma(x-x')\right>
+\frac{1}{2}\left<x,i\cR x'\right>\right\}.
\ee
\end{lemma}
{\it Proof.} Let $f\in C_0^\infty(\RR^n)$. Then by using
(\ref{612igaxm})
we have
\bea
&&
\hspace{-1cm}
(Kf)(x) 
= 
(4\pi t)^{-n/2}(\det\gamma)^{1/2}\int\limits_{\RR^n} d\xi
\exp\left\{-\frac{1}{4t}\left<\xi, \gamma\xi\right>
-\frac{1}{2}\left<\xi,i\cR x\right>\right\}
f(x+\xi)
\nonumber
\\
&=&
(4\pi t)^{-n/2}(\det\gamma)^{1/2}
\int\limits_{\RR^n} d\xi
\exp\left\{-\frac{1}{4t}\left<(\xi-x), \gamma(\xi-x)\right>
-\frac{1}{2}\left<\xi,i\cR x\right>\right\}
f(\xi)\,.
\nonumber\\
\eea
This integral obviously 
converges for any smooth function of compact support.
The integral kernel of the operator $K$ is obtained by
acting on the delta-function,
$K(t;x,x')=K\delta(x-x')$.
$\Box$

By expanding this operator in the Taylor series 
and using (\ref{229igan}) it is easy
to obtain
\be
\left<\exp\left<\xi,\nabla\right>\right>_t
=\sum_{k=0}^\infty \frac{t^{k}}{k!} 
\gamma^{(i_1i_2}\cdots\gamma^{i_{2k-1}i_{2k})}
\nabla_{(i_1}\cdots\nabla_{i_{2k)}}.
\ee
Notice that in the trivial case when $\cR=0$,
that is, when the operators $\nabla_i$ commute,
we have
$\left<\exp\left<\xi,\nabla\right>\right>_t=\exp(t\Delta_\gamma)$,
where $\Delta_\gamma=\gamma^{ij}\nabla_i\nabla_j$.

In the following lemma we compute this Gaussian average
for non-commuting operators $\nabla_i$
forming the Lie algebra   (\ref{nabla-comm})
following
\cite{avramidi93,avramidi15}.
Let $G^{-1}(t)=(G^{ij})$ be a symmetric matrix defined 
by
\be 
G^{-1}(t) = \Phi(t\gamma^{-1}i\cR)\gamma^{-1}
=\frac{\tanh^{-1}(t\gamma^{-1}i\cR)}{t\gamma^{-1}i\cR}\gamma^{-1}
\label{833viax0}
\ee
and $\Delta_{G(t)}$ be an operator defined by
\be
\Delta_{G(t)}=\left<\nabla,G^{-1}(t)\nabla\right>
=G^{ij}(t)\nabla_i\nabla_j.
\label{833viax}
\ee
For the future reference we explore the behavior of the metric $G(t)$
for small $t$,
\be
G(t)=\gamma
-\frac{1}{3}t^2i\cR\gamma^{-1}i\cR
+O(t^4).
\ee

\begin{lemma}
\label{lemma3}
For sufficiently small $t>0$ 
\be
\left<\exp\left<\xi,\nabla\right>\right>_t=
\det\left(I+t\gamma^{-1}i\cR\right)^{-1/2}
\exp\left(t\Delta_{G(t)}\right)\,.
\label{833viaxz}
\ee
\end{lemma}
{\it Proof.}
Let $I(t)=\left<\exp\left<\xi,\nabla\right>\right>_t$.
We compute the derivative of the function
$I(t)$ by using (\ref{222cbxt})
\be
\partial_t I(t)=\frac{1}{2t}\nabla_i\left<\xi^i\exp\left<\xi,\nabla\right>\right>_{t}.
\label{222cbx}
\ee
Next, by using (\ref{835viax}) we get
\be
\left<\xi^i\exp\left<\xi,\nabla\right>\right>_{t}
=2t\gamma^{ik}\left<\frac{\partial}{\partial \xi^k}\exp\left<\xi,\nabla\right>\right>_{t}
\label{234iga}
\ee
Further, by using eqs. (\ref{835viax}) and (\ref{829viaz})
we have
\be
\Gamma_{ij}\gamma^{jk}\left<\frac{\partial}{\partial \xi^k}\exp\left<\xi,\nabla\right>\right>_{t}
=\nabla_i\left<\exp\left<\xi,\nabla\right>\right>_{t},
\ee
where $\Gamma=(\Gamma_{ij})$ is the matrix
\be
\Gamma=\gamma+ti\cR;
\ee
and, therefore,
\be
\gamma^{jk}\left<\frac{\partial}{\partial \xi^k}\exp\left<\xi,\nabla\right>\right>_{t}
=\Gamma^{ji}\nabla_i\left<\exp\left<\xi,\nabla\right>\right>_{t},
\label{838viax}
\ee
where $\Gamma^{-1}=(\Gamma^{ij})$ is the inverse of the matrix 
$\Gamma=(\Gamma_{ij})$.
By substituting (\ref{838viax}) in (\ref{234iga}) we get
\be
\left<\xi^j\exp\left<\xi,\nabla\right>\right>_{t}
=2t\Gamma^{ji}\nabla_i\left<\exp\left<\xi,\nabla\right>\right>_{t},
\label{838iga}
\ee
and, finally, by using this equation in (\ref{222cbx})
we obtain a differential equation
for the function $I(t)$,
\be
\partial_t I(t)=L(t) I(t),
\label{841viax}
\ee
where
\be
L=\Gamma^{ij}\nabla_i\nabla_j,
\ee
with the initial condition $I(0)=I$.

We decompose the matrix $\Gamma^{-1}=(\Gamma^{ij})$ in the symmetric and the
anti-symmetric parts and use the commutator of the operators $\nabla_i$
to get
\be
L(t) = \Delta_{S(t)}+P(t),
\ee
where
\bea
P(t) &=& -\frac{1}{2}\tr\left\{\left(I+t\gamma^{-1}i\cR\right)^{-1}\gamma^{-1}i\cR\right\},
\\
\Delta_{S(t)} &=& S^{ij}(t)\nabla_i\nabla_j,
\eea
with $S^{-1}=(S^{ij})$ being a symmetric matrix 
defined by
\be
S^{-1}=\frac{1}{2}\left\{
\left(I+t\gamma^{-1}i\cR\right)^{-1}
+\left(I-t\gamma^{-1}i\cR\right)^{-1}\right\}\gamma^{-1}.
\ee

Next, by using the fact that the matrix $S^{-1}(t)$
at different times satisfies the equation
\be
S^{-1}(t)\cR S^{-1}(s)
= S^{-1}(s)\cR S^{-1}(t)
\ee
we can use Lemma \ref{lemma1} to show that
the operators $\Delta_{S(t)}$
and, therefore, the operators
$L(t)$, at different times
commute \cite{avramidi93},
\be
[L(t),L(s)]=0.
\ee

This enables one to 
solve the differential equation (\ref{841viax})
with the initial condition $I(0)=I$ to obtain
\be
I(t)=\exp\left\{\int_0^t d\tau\; L(\tau)\right\}
=\exp\left\{t\Delta_{G(t)}+M(t)\right\},
\label{845viaxz}
\ee
where
\be
M(t) = -\frac{1}{2}\tr \log\left(I+t\gamma^{-1}i\cR\right),
\ee
and 
$\Delta_{G(t)}$ is given by
(\ref{833viax})
with
\be 
G^{-1}(t)=\frac{\tanh^{-1}\left(t\gamma^{-1}i\cR\right)}{t\gamma^{-1}i\cR}
\gamma^{-1}.
\label{833viax1}
\ee
Thus, we obtain from (\ref{845viaxz})
\be
I(t)=\det\left(I+t\gamma^{-1}i\cR\right)^{-1/2}
\exp\left(t\Delta_{G(t)}\right).
\label{845viax}
\ee
$\Box$

By using Lemma \ref{lemma3} we immediately obtain
the following corollary.
\begin{corollary}
\label{corollary3}
Let $A_i$ be a vector commuting with the operators $\nabla_j$. 
Then for sufficiently small $t>0$
\bea
&&\int\limits_{\RR^n} d\xi
\exp\left\{-\frac{1}{4t}\left<\xi, \gamma\xi\right>
+\left<A,\xi\right>\right\}
\exp\left<\xi,\nabla\right>
=(4\pi t)^{n/2}
\det(\gamma+ti\cR)^{-1/2}
\nonumber
\\
&&\qquad \times
\exp\left\{t\Delta_{G(t)}
+2t\left<G^{-1}(t)A,\nabla\right>
+t\left<A,G^{-1}(t)A\right>\right\}\,.
\label{833viaq}
\eea
\end{corollary}
{\it Proof.}
We notice that the operators
$\nabla_i+A_i$ form the same Lie algebra
as the operators $\nabla_i$. Therefore,
by replacing $\nabla_i\mapsto \nabla_i+A_i$
we obtain (\ref{833viaq}).
$\Box$

This lemma enables one to prove the following theorem
for the heat semigroup \cite{avramidi93,avramidi15}.
Notice that, although Lemma \ref{lemma3} is valid, strictly speaking,
only for small $t$, this theorem 
holds for any $t$.
Let  $g^{-1}=(g^{ij})$ be a symmetric positive matrix,
and 
$\Delta_g=g^{ij}\nabla_i\nabla_j$.
Let $D(t)=(D_{ij})$
be a symmetric matrix defined by
\bea
D(t) &=& \frac{1}{t}g\Psi(tg^{-1}i\cR)=i\cR\coth\left(tg^{-1}i\cR\right)\,
\eea
and $\Omega(t)$ be a function defined by
\bea
\Omega(t) &=& 
\det(D(t)+i\cR)^{1/2}
=
\det \left(g^{-1}\frac{\sinh(tg^{-1}i\cR)}{g^{-1}i\cR}\right)^{-1/2}.
\label{257igan}
\eea
For the future reference we explore
the Taylor expansion of these objects for small $t$ 
\bea
D(t) &=& \frac{1}{t}\left\{g+\frac{1}{3}t^2i\cR g^{-1}i\cR+O(t^4)\right\},
\\
\Omega(t) &=& t^{-n/2}(\det g)^{1/2}\left\{1
-\frac{1}{12}t^2\tr \left(g^{-1}i\cR g^{-1}i\cR\right)
+O(t^4)\right\}.
\eea

\subsection{Proof of Theorem \ref{theorem2viaz}}

Let $J(t)$ be the right hand side of eq. (\ref{858viax}).
By using Lemma \ref{lemma3} to compute the integral
over $\xi$ (with $\gamma=tD(t)$) we obtain from (\ref{833viaxz})
or (\ref{833viaq})
\bea
J(t) &=& \Omega(t) 
\det\left(D(t)+i\cR\right)^{-1/2}
\exp\left(t\Delta_{G(t)}\right),
\label{858viaz}
\eea
where the matrix $G^{-1}(t)$ is given by
\be
G^{-1}(t)=\frac{1}{t}\Phi\left(D^{-1}(t)i\cR\right)D^{-1}(t).
\ee
It is easy to see that
\be
D^{-1}(t)i\cR=\tanh\left(tg^{-1}i\cR\right),
\ee
therefore, by using (\ref{222igaz}) we get 
\be
G^{-1}(t)
=\frac{1}{t}\Psi(tg^{-1}i\cR)D^{-1}(t)=g^{-1},
\ee
and $\Delta_{G(t)}=\Delta_g$.
Also, we notice that
\bea
\det\left(D(t)+i\cR\right)
&=&\det\left\{(D+i\cR)(D-i\cR)\right\}^{1/2}
\nonumber\\
&=&
\det\left(\frac{i\cR}{\sinh(tg^{-1}i\cR)}\right)
=\Omega^2(t).
\label{247cbx}
\eea
Therefore, by using eq. (\ref{247cbx})
in (\ref{858viaz})
we obtain $J(t)=\exp(t\Delta_g)$.
$\Box$

By using Lemma \ref{lemma1viaz} and Theorem \ref{theorem2viaz}
one can prove the following corollary.
\begin{corollary}
Let $T(t)$ be the matrix defined by
\be
T(t)=D(t)+i\cR.
\label{268igan}
\ee
Then
\bea
\nabla_k\exp(t\Delta_g)&=&
\frac{1}{2}(4\pi )^{-n/2}\Omega(t) 
\int\limits_{\RR^n} d\xi
\exp\left\{-\frac{1}{4}\left<\xi, D(t)\xi\right>\right\}
T_{kj}(t)\xi^j\exp\left<\xi,\nabla\right>\,,
\nonumber\\
\\
\exp(t\Delta_g)\nabla_k&=&
\frac{1}{2}(4\pi )^{-n/2}\Omega(t)
\int\limits_{\RR^n} d\xi
\exp\left\{-\frac{1}{4}\left<\xi, D(t)\xi\right>\right\}
T_{jk}(t)\xi^j\exp\left<\xi,\nabla\right>\,.
\nonumber\\
\eea
\end{corollary}
{\it Proof.} This is proved by using the Lemma \ref{lemma1viaz} and integrating by parts.
$\Box$


\section{Product of Semigroups}
\setcounter{equation}0

We consider the set of anti-self-adjoint
operators $(\nabla_1^+,\dots,\nabla_n^+,
\nabla^-_1, \dots,\nabla_n^-, i)$ 
acting on the space $C^\infty_0(\RR^n)$ of smooth functions of compact support
forming the Lie algebra
\bea
[\nabla^+_i,\nabla^+_j] &=& i\cR^+_{ij},
\\{}
[\nabla^-_i,\nabla^-_j] &=& i\cR^-_{ij},
\\{}
[\nabla^+_i,\nabla^-_j] &=& i\cR_{ij},
\label{843viaz}
\eea
where
\be
\cR_{ij}=\frac{1}{2}\left(\cR^+_{ij}+\cR^-_{ij}\right).
\label{84igax}
\ee
This algebra can be written in a more compact form by
introducing the operators $(\tilde\cD_A)$, $A=1,\dots,2n$,
by
\bea
\tilde\cD_1 &=& \nabla^+_1, \;\dots,\; \tilde\cD_n=\nabla^+_{n},\;
\\
\tilde\cD_{n+1} &=& \nabla^-_{1},\;\dots,\;\tilde\cD_{2n}=\nabla^-_{n}.
\eea
We will use the convention that  
the capital Latin indices run over $1,\dots, 2n$.
The operators $\tilde\cD_A$ form the algebra
\bea
[\tilde\cD_A, \tilde\cD_B] &=& i\tilde\cF_{AB},
\label{843viazxt}
\eea
where $\tilde\cF=(\tilde\cF_{AB})$ is a $2n\times 2n$ 
anti-symmetric matrix 
defined by
\bea
\tilde\cF
&=&\left(
\begin{array}{cc}
\cR^+ & \cR\\
\cR & \cR^-
\end{array}
\right).
\eea

The same algebra can be also rewritten in more convenient form
by explicitly exhibiting an Abelian subalgebra.
Let
\bea
\nabla_i&=&\frac{1}{2}(\nabla^+_i+\nabla^-_i),
\label{351via}
\\
X_i&=&\nabla^+_i-\nabla^-_i,
\label{352viam}
\eea
so that
\be
\nabla^\pm_i=\nabla_i\pm \frac{1}{2}X_i.
\ee
These operators form the algebra
\bea
[\nabla_i,\nabla_j] &=& i\cR_{ij},
\\{}
[\nabla_i,X_j] &=& iF_{ij},
\\{}
[X_i,X_j] &=& 0,
\label{843viazx}
\eea
where
\be
F_{ij}=\frac{1}{2}\left(\cR^+_{ij}-\cR^-_{ij}\right).
\label{815igax}
\ee
Let us also define
$(\cD_A)=(\cD_1,\dots,\cD_n,
\cD_{n+1},\dots,\cD_{2n})$,
\bea
\cD_1 &=& \nabla_1, \;\dots,\; \cD_n=\nabla_{n},\;
\\
\cD_{n+1} &=& X_1,\;\dots,\;\cD_{2n}=X_{n},
\eea
These operators satisfy the commutation relations
\bea
[\cD_A, \cD_B] &=& i\cF_{AB},
\label{843igaz}
\eea
where $\cF=(\cF_{AB})$ is a $2n\times 2n$ 
anti-symmetric matrix 
defined by
\bea
\cF
&=&\left(
\begin{array}{cc}
\cR & F\\
F & 0
\end{array}
\right).
\eea

These bases are related by
\be
\tilde\cD=\Lambda\cD,
\ee
where $\Lambda$ is a $2n\times 2n$ matrix
defined by
\be
\Lambda
=
\left(
\begin{array}{cc}
I & \frac{1}{2}I\\
I & -\frac{1}{2}I
\end{array}
\right)
\label{2122igap}
\ee
and, therefore,
\be
\tilde\cF=\Lambda\cF\Lambda^T.
\ee


Let $g^{ij}_\pm$ be two positive symmetric matrices
defining the Laplacians
\bea
\Delta_\pm &=& 
\left<\nabla_\pm,g_\pm^{-1}\nabla_\pm\right>
=g_\pm^{ij}\nabla^\pm_i\nabla^\pm_j.
\eea
Our goal is the computation of the convolution
\be
U(t,s)=\exp(t\Delta_+)\exp(s\Delta_-).
\ee

We write the matrices $g_\pm^{-1}$ in the form
$g_\pm^{-1}=\omega_\pm\omega_\pm^T$. Then we have
$\Delta_\pm=\left<\nabla'_\pm,\nabla'_\pm\right>$,
with $\nabla'_\pm=\omega^T_\pm\nabla_\pm$
satisfying the commutation relations with the
curvatures $\cR'_\pm=\omega^T_\pm\cR_\pm\omega_\pm$.
Further, we can subject the operators $\nabla'_\pm$ to
 orthogonal transformations
$\tilde \nabla_\pm=O_\pm\nabla'_\pm=O_\pm\omega^T_\pm\nabla_\pm$
with  orthogonal matrices $O_\pm$ without changing the
operators $\Delta_\pm$. Then 
$\Delta_\pm=\left<\tilde\nabla_\pm,\tilde\nabla_\pm\right>$
and the transformed curvatures are
$\tilde\cR_\pm=O_\pm\omega^T_\pm\cR_\pm\omega_\pm O^T_\pm$;
the matrices $O_\pm$ can be chosen so that the curvatures
$\tilde \cR_\pm$ have the canonical form (\ref{311igaz}).
Thus, without loss of generality we can assume that
the matrices $g_\pm$ are equal to the unit matrix, $g_\pm=I$,
and the matrices $\cR_\pm$ have the canonical form.

\subsection{Method I}



In the following we use the functions $\Phi(z)$ 
and $\Psi(z)$
defined by (\ref{219iga}) and (\ref{220iga})
and the matrix $\Lambda$ defined by
(\ref{2122igap}).
Let $D_\pm $ be two
positive symmetric matrices defined  by
\bea
D_\pm(t) &=& \frac{1}{t}g_\pm\Psi(tg_\pm^{-1}i\cR_\pm)
\label{725igaz}
\,.
\eea
Let $\tilde Q(t,s)=(\tilde Q_{AB})$ and
$Q(t,s)=(Q_{AB})$
be the 
$2n\times 2n$ symmetric matrices
defined by
\bea
\tilde Q(t,s)
&=&\left(
\begin{array}{cc}
D_+(t) & -i\cR\\
i\cR & D_-(s)
\end{array}
\right),
\\
Q(t,s)
&=&\Lambda^{-1}\tilde Q(t,s)\Lambda^{-1}{}^T
\nonumber\\
&=&
\left(
\begin{array}{cc}
\frac{1}{4}(D_++D_-) & \frac{1}{2}(D_+-D_-+2i\cR)\\
\frac{1}{2}(D_+-D_--2i\cR) & D_++D_-
\end{array}
\right),
\eea
and
$\tilde\cG^{-1}(t,s)=(\tilde\cG^{AB})$ and 
$\cG^{-1}(t,s)=(\tilde\cG^{AB})$
be the 
$2n\times 2n$ symmetric matrices
defined by
\bea
\tilde\cG^{-1}(t,s)
&=& \Phi(\tilde Q^{-1}(t,s)i\tilde\cF)\tilde Q^{-1}(t,s),
\\
\cG^{-1}(t,s)&=& \Lambda^T\tilde \cG^{-1}(t,s)\Lambda
\nonumber\\
&=& \Phi(Q^{-1}(t,s)i\cF) Q^{-1}(t,s).
\eea

Let $\cH(t,s)$ be the operator defined by
\bea
\cH(t,s) &=& \left<\cD,\cG^{-1}(t,s)\cD\right>;
\eea
it is easy to see that it is also equal to
\bea
\cH(t,s)=\left<\tilde\cD,\tilde\cG^{-1}(t,s)\tilde\cD\right>.
\label{831igax}
\eea
We can write the operator $\cH$ in a more explicit form.
We represent the metric $\cG^{-1}$ in the block diagonal form
\be
\cG^{-1}=
\left(
\begin{array}{cc}
G^{-1} & Y\\
Y^T & M
\end{array}
\right),
\ee
where $G$ and $M$ are symmetric matrices.
Then the operator $\cH$ takes the form
\bea
\cH &=& \left<\nabla,G^{-1}\nabla\right>
+\left<\nabla, Y X\right>
+\left<X, Y^T\nabla\right>
+\left<X,MX\right>,
\nonumber\\
&=&
\left<\hat\nabla,G^{-1}\hat\nabla\right>
+\left<X,V X\right>,
\eea
where
\bea
\hat \nabla&=& \nabla+GYX= \frac{1}{2}(I+2GY)\nabla_+
+\frac{1}{2}(I-2GY)\nabla_-,
\\
V &=& M-Y^TGY.
\eea


\subsubsection{Proof of Theorem \ref{theorem2in}}

By using Theorem \ref{theorem2viaz} we have
\bea
U(t,s)&=&
(4\pi)^{-n}
\Omega_+(t)\Omega_-(s)
\nonumber\\
&&\times
\int\limits_{\RR^{2n}} d\xi_+\,d\xi_-
\exp\left\{-\frac{1}{4}\left<\xi_+,D_+(t)\xi_+\right>
-\frac{1}{4}\left<\xi_-,D_-(s)\xi_-\right>
\right\}
\nonumber\\
&&\times
\exp\left<\xi_+,\nabla^+\right>
\exp\left<\xi_-,\nabla^-\right>\,,
\label{299igan}
\eea
where the functions $\Omega_\pm$ are defined by (\ref{257igan}).
By using the special case of the Campbell-Hausdorff formula
(\ref{285igam})
and the commutators (\ref{843viaz})
we get
\bea
U&=&
(4\pi)^{-n}
\Omega_+\Omega_-
\nonumber\\
&&\times
\int\limits_{\RR^{2n}} d\xi_+\,d\xi_-
\exp\left\{-\frac{1}{4}\left<\xi_+,D_+\xi_+\right>
-\frac{1}{4}\left<\xi_-,D_-\xi_-\right>
+\frac{1}{2}\left<\xi_+,i\cR\xi_-\right>\right\}
\nonumber\\
&&\times
\exp\left\{\left<\xi_+,\nabla^+\right>
+\left<\xi_-,\nabla^-\right>\right\}.
\label{358viac}
\eea

We introduce new integration variables $(\eta^A)
=(\xi_+^1,\dots,\xi_+^n,\xi_-^1,\dots,\xi_-^n)$
and use the operators $\tilde\cD_A$ to rewrite 
this in the form
\bea
U&=&
(4\pi)^{-n}
\Omega_+\Omega_-
\int\limits_{\RR^{2n}} d\eta
\exp\left\{-\frac{1}{4}\left<\eta, \tilde Q\eta\right>
\right\}
\exp\left<\eta,\tilde\cD\right>.
\label{358igan}
\eea
This integral can be computed by using Corollary
\ref{corollary3} to obtain
\be
U(t,s)=\frac{\Omega_+(t)\Omega_-(s)}{\tilde\Omega(t,s)}
\exp\left<\tilde\cD, \tilde\cG^{-1}(t,s)\tilde\cD\right>,
\ee
where
\be
\tilde\Omega(t,s)=\det\left(\tilde D(t,s)+i\tilde\cF\right)^{1/2}.
\ee
Finally, we compute the determinant
\be
\det\left(\tilde D(t,s)+i\tilde\cF\right)
=\det(D_++i\cR_+)\det(D_-+i\cR_-)=\Omega_+^2(t)\Omega_-^2(s);
\ee
therefore, $\tilde\Omega(t,s)=\Omega_+(t)\Omega_-(s)$,
which proves the theorem by taking into account (\ref{831igax}).
$\Box$

\subsubsection{Examples}

Notice that for small $t,s$ the metric factorizes
\be
\tilde \cG^{-1}
=\left(
\begin{array}{cc}
tI & 0\\
0 & s I
\end{array}
\right)+\cdots,
\ee
so that
\be
\cH= t\Delta_++s\Delta_-+\cdots.
\ee

Let us study another particular case, when the operators are equal,
that is,  $\nabla_+=\nabla_-=\nabla$, $\cR_+=\cR_-=\cR$.
Then we should have obviously
\be
\tilde\cH=(t+s)\Delta.
\ee
Then there is only one matrix 
in the problem, $\cR$. Therefore, all matrices commute.
To simplify notation let $x=ti\cR$ and $y=si\cR$
and
\be
a=\coth x, \qquad b=\coth y.
\ee 
Then
\bea
i\tilde\cF
&=&\frac{x+y}{t+s}\left(
\begin{array}{cc}
1 & 1\\
1 & 1
\end{array}
\right)
\qquad
\mbox{and}\qquad
\tilde Q
=\frac{x+y}{t+s}\left(
\begin{array}{cc}
a & -1\\
1 & b
\end{array}
\right).
\eea
The inverse of the matrix $\tilde D$
has the form
\bea
\tilde Q^{-1}
&=&\frac{(t+s)}{(x+y)}\frac{1}{(ab+1)}
\left(
\begin{array}{cc}
b & 1\\
-1 & a
\end{array}
\right);
\eea
therefore,
\be
\tilde Q^{-1}i\tilde\cF
=c\Pi,
\ee
where 
and
$
c=\frac{a+b}{ab+1}=\tanh(x+y)
$
and
\be
\Pi
=\frac{1}{a+b}
\left(
\begin{array}{cc}
b+1 & b+1\\
a-1 & a-1
\end{array}
\right).
\ee
It is easy to see that the matrix $\Pi$ is idempotent
\be
\Pi^2=\Pi,
\ee
and, therefore, for any analytic function of $\Pi$
\be
f(t\Pi)=f(0)(I-\Pi)+f(t)\Pi.
\ee
Therefore, 
by using (\ref{222igaz}) we have
\bea
\Phi(\tilde Q^{-1}i\tilde\cF)
&=&
I+\left[\frac{(x+y)}{c}-1\right]\Pi.
\eea
Further, we compute
the metric
\bea
\tilde\cG^{-1} &=& 
\Phi(\tilde Q^{-1}i\tilde\cF)\tilde Q^{-1}
\nonumber\\
&=&(t+s)
\Biggl\{
\frac{1}{(a+b)^2}
\left(
\begin{array}{cc}
b^2-1 & (a+1)(b+1)\\
(a-1)(b-1) & a^2-1
\end{array}
\right)
\nonumber\\
&&
+
\frac{1}{(x+y)(a+b)}
\left(
\begin{array}{cc}
1 & -1\\
-1 & 1
\end{array}
\right)
\Biggr\}.
\eea
By using this metric 
(and noticing that the sum of all its elements is equal
to $(t+s)$)
it is easy to show that indeed
the operator $\cH$ takes the form
\be
\cH
=\left<\tilde\cD,\tilde G^{-1}\tilde\cD\right>
=(t+s)\Delta.
\ee


\subsection{Method II}

%

Finally, we present a method that will be useful
to compute the kernel of the product
$\exp(t\Delta_+)$ $\exp(s\Delta_-)$.
We recall the definition of the matrices  
$T_\pm(t)=D_\pm(t)+i\cR_-$ by (\ref{268igan}). 
Also, we define two
matrices
\bea
D(t,s)&=& D_+(t)+D_-(s),
\label{762igaz}
\\
Z(t,s) &=& D_+(t)-D_-(s)-2i\cR_-,
\label{491viac}
\eea
and a function $\Omega(t,s)$ by
\be
\Omega(t,s) = \Omega_+(t)\Omega_-(s)\det D^{-1/2}(t,s).
\label{2128igan}
\ee
We also define a symmetric matrix $H$ by
\be
H(t,s) = \frac{1}{4}\left(D(t,s)-Z^T(t,s)D^{-1}(t,s)Z(t,s)\right).
\label{859igax}
\ee
By rewriting the matrix $Z$ in the form
\be
Z=D-2T_-
\ee
the matrix $H$ takes the form
\bea
H(t,s) &=& D_-(s)-T_-^T(s)D^{-1}(t,s)T_-(s)
\label{2142igab}
\eea
and by writing the matrix $Z$ in the form
\be
Z=-D+2T^T_++4iF
\label{2130igam}
\ee
we can represent the matrix $H$ in terms of the matrices
$D_+$ and $T_+$ by
\bea
H(t,s)
&=&
D_+(t) -T_+(t)D^{-1}(t,s)T^T_+(t)
\label{2131igam}\\
&&
-2T_+(t)D^{-1}(t,s)iF
+2iFD^{-1}(t,s)T^T_+(t)
+4iFD^{-1}(t,s)iF.
\nonumber
\eea
We exhibit the behavior of these objects for small $t,s$,
\bea
D(t,s)&=& \left(\frac{1}{t}+\frac{1}{s}\right)I+\cdots,
\\
Z(t,s)&=& \left(\frac{1}{t}-\frac{1}{s}\right)I+\cdots,
\\
\Omega(t,s)&=& (t+s)^{-n/2}+\cdots,
\\
H(t,s) &=&
\left(s+t\right)^{-1}I+\cdots
\eea

\subsubsection{Proof of Theorem \ref{theorem3in}}

We change the integration variables in (\ref{358viac}) by
\be
\xi_+=\frac{1}{2}\alpha+\beta, \qquad
\xi_-=\frac{1}{2}\alpha-\beta,
\ee
so that
\be
\alpha=\xi_++\xi_-, \qquad
\beta=\frac{1}{2}(\xi_+-\xi_-),
\ee
to obtain
\bea
U(t,s)&=&
(4\pi)^{-n}
\Omega_+\Omega_-
\int\limits_{\RR^{2n}} d\alpha\,d\beta
\exp\left\{-\frac{1}{16}\left<\alpha,D\alpha\right>
-\frac{1}{4}\left<\beta,D\beta\right>
\right\}
\nonumber\\
&&\times
\exp\left\{
-\frac{1}{4}\left<\beta,(D_+-D_--2i\cR)\alpha\right>\right\}
\exp\left\{\left<\alpha,\nabla\right>
+\left<\beta,X\right>\right\}.
\label{2138igab}
\eea
%
By using the Campbell-Hausdorff formula (\ref{285igam}) we obtain
\bea
\exp\left\{\left<\alpha,\nabla\right> 
+\left<\beta,X\right>
\right\}
=
\exp\left\{-\frac{1}{2}\left<\beta,iF\alpha\right>\right\}
\exp\left<\beta, X\right>
\exp\left<\alpha, \nabla\right>.
\eea
Therefore, eq. (\ref{2138igab}) takes the form
\bea
U&=&
(4\pi)^{-n}
\Omega_+\Omega_-
\int\limits_{\RR^{2n}} d\alpha d\beta
\exp\left\{-\frac{1}{4}\left<\beta,D\beta\right>
-\frac{1}{4}\left<\beta,Z\alpha\right>
+\left<\beta, X\right>
\right\}
\nonumber\\
&&\times
\exp\left\{-\frac{1}{16}\left<\alpha,D\alpha\right>\right\}
\exp\left<\alpha, \nabla\right>,
\eea
with the matrices $D$ and $Z$ defined by (\ref{762igaz}) and (\ref{491viac}).
The integral over $\beta$ can be computed 
by using Corollary \ref{corollary3}
to obtain (\ref{490viaxc})
with the matrix $H$ in the form
(\ref{859igax}).
$\Box$


\section{Convolution of Heat Kernels}
\setcounter{equation}0


We work with operators 
\be
\nabla_i^\pm=\partial_i-\frac{1}{2}i\cR^\pm_{ij}x^j,
\ee
acting on smooth functions in $\RR^n$.
These operators form the Lie algebra (\ref{843viaz}).
The operators $\nabla_i$ and $X_i$ defined in
(\ref{351via}) and (\ref{352viam}) then take the form
\bea
\nabla_i &=& \partial_i-\frac{1}{2}i\cR_{ij}x^j,
\\
X_i &=& -iF_{ij}x^j,
\label{495viaxc}
\eea
with the matrices $\cR_{ij}$ and $F_{ij}$ 
defined in (\ref{84igax}) and (\ref{815igax}).
The corresponding Laplacians are
$\Delta_\pm=g_\pm^{ij}\nabla^\pm_i\nabla^\pm_j$. 


Recall the definition of the matrices $D_\pm(t)$ 
and $T_\pm(t)$ by (\ref{725igaz})
and (\ref{268igan}). Also, notice that the functions
$\Omega_\pm(t)$ defined by (\ref{257igan})
are determined by the determinant of the
matrices $T_\pm$
\be
\Omega_\pm(t)=\det T_\pm^{1/2}.
\ee
Let $S_\pm$ be functions defined by
\bea
S_\pm(t;x,x') &=& 
\frac{1}{4}\left<(x-x'),D_\pm(t) (x-x')\right>
-\frac{1}{2}\left<x,i\cR_\pm x'\right>
\nonumber\\
&=&
\frac{1}{4}\left<x,D_\pm(t)x\right>
+\frac{1}{4}\left<x',D_\pm(t)x'\right>
-\frac{1}{2}\left<x, T_\pm(t)x'\right>,
\eea
Notice that the matrix of second mixed partial derivatives
of the function $S_\pm$ is proportional to the matrix $T_\pm$
\be
S^\pm_{xx'}=-\frac{1}{2}T_\pm(t),
\ee
and, therefore,
\be
\Omega_\pm(t)=\det\left(-2S^\pm_{xx'}\right)^{1/2}.
\label{86igaz}
\ee


\begin{lemma}
The heat kernel of the operator $\Delta_\pm$ is
\be
U_\pm(t;x,x')
=\det\left(-\frac{S^\pm_{xx'}}{2\pi}\right)^{1/2}\exp\left\{-S_\pm(t;x,x')\right\}.
\ee
\end{lemma}
{\it Proof.}
The heat kernel is given by
$
U_\pm(t;x,x')=\exp(t\Delta_\pm)
\delta(x-x').
$
We use eq. (\ref{858viax}) to compute it,
\bea
U_\pm(t;x,x')
&=&
(4\pi)^{-n/2}\Omega_\pm(t)
\\
&&\times\int\limits_{\RR^n} d\xi_\pm
\exp\left\{-\frac{1}{4}\left<\xi_\pm,D_\pm(t) \xi_\pm\right>
\right\}
\exp\left<\xi_\pm,\nabla_\pm\right>\delta(x-x').
\nonumber
\eea
We notice that the operators
$\left<\xi_\pm,\partial\right>$ and $\left<\xi_\pm,\cR_\pm x\right>$
commute; therefore,
we have
\be
\exp\left<\xi_\pm,\nabla_\pm\right>
=
\exp\left\{-\frac{1}{2}\left<\xi_\pm,i\cR_\pm x\right>\right\}
\exp\left<\xi_\pm,\partial\right>
\ee
and, hence,
\bea
\exp\left<\xi_\pm,\nabla_\pm\right>\delta(x-x')
&=&
\exp\left\{-\frac{1}{2}\left<\xi_\pm,i\cR_\pm x\right>\right\}
\delta(x-x'+\xi_\pm).
\eea
This immediately gives the heat kernel
\be
U_\pm(t;x,x')
=(4\pi)^{-n/2}\Omega_\pm(t)\exp\left\{-S_\pm(t;x,x')\right\},
\ee
which proves the lemma by taking into account (\ref{86igaz}).
$\Box$


We proceed as follows.
Let $S$ be a function defined by
\bea
S(t,s;x,x')&=&
\frac{1}{4}\left<x, A_+(t,s)x\right>
+\frac{1}{4}\left<x', A_-(t,s) x'\right>
-\frac{1}{2}\left<x,B(t,s)x'\right>,
\label{316igan}
\eea
where
\bea
A_+&=& H
+Z^TD^{-1}iF-iFD^{-1}Z
+4iFD^{-1}iF,
\label{317igan}\\
A_-&=& H,
\\
B&=& 
H-iFD^{-1}Z+i\cR.
\eea
By using the definition of the matrix $H$, (\ref{2142igab}) and (\ref{2131igam}),
and of the matrix $Z$, (\ref{491viac}) and (\ref{2130igam}),
one can show that
\bea
A_+(t,s) &=& D_+(t)-T_+^T(t)D^{-1}(t,s)T_+(t),
\\
A_-(t,s) &=& D_-(s)-T_-^T(s)D^{-1}(t,s)T_-(s),
\\
B(t,s)&=& T_+(t)D^{-1}(t,s)T_-(s).
\eea

\subsection{Proof of Theorem \ref{theorem4in}}

The kernel of the product of the semigroups
$
U(t,s;x,x')
$ 
can be computed by using the equation (\ref{490viaxc}).
Since the operators
$\left<\alpha,\partial\right>$ and $\left<\alpha,\cR x\right>$
commute
we have
\bea
\exp\left<\alpha,\nabla\right>\delta(x-x')
&=&\exp\left\{-\frac{1}{2}\left<\alpha,i\cR x\right>\right\}
\delta(x-x'+\alpha).
\eea
Therefore, we immediately obtain from (\ref{490viaxc})
\be
U(t,s;x,x')
=
(4\pi)^{-n/2}\Omega(t,s)
\exp\left\{-S(t,s;x,x')\right\},
\label{519viaz}
\ee
where $\Omega(t,s)$ is defined by (\ref{2128igan})
 and
$S$ is a function given by
\bea
S(t,s;x,x') &=&
\frac{1}{4}\left<(x-x'), H(x-x')\right>
-\frac{1}{2}\left<(x-x'),Z^TD^{-1}X\right>
\nonumber\\
&&
-\left<X,D^{-1}X\right>
-\frac{1}{2}\left<x,i\cR x'\right>.
\eea
with $H$ and $Z$ being the matrices defined by (\ref{2142igab}) and (\ref{491viac}).
It is not difficult to simplify this to the form
(\ref{316igan}).
Further, we see that
\be
\Omega(t,s)=\frac{\det \Omega_+(t)\Omega_-(s)}{\det D^{1/2}(t,s)}
=\det B^{1/2}=\det\left(-2S_{xx'}(t,s)\right)^{1/2},
\ee
which gives finally (\ref{820iganm}).
$\Box$





As an independent check we compute the convolution of the two heat kernels
directly by computing the integral
\be
U(t,s;x,x')=\int_{\RR^n}dz\; U_+(t;x,z)U_-(s;z,x').
\ee
This integral is Gaussian and can be easily computed to obtain
\be
U(t,s;x,x')
=
(4\pi)^{-n/2}\Omega(t,s)
\exp\left\{-S(t,s;x,x')\right\},
\label{519igan}
\ee
where $\Omega(t,s)$ is defined by (\ref{2128igan}) and
$S$ is a function given by
\be
S(t,s;x,x')=\frac{1}{4}\left<x,D_+(t)x\right>
+\frac{1}{4}\left<x'D_-(s)x'\right>
-\frac{1}{4}\left<y,D^{-1}(t,s)y\right>,
\ee
with
\be
y=T_+^T(t)x+T_-(s)x'.
\ee
It is easy to see that
it is indeed equal to (\ref{316igan}).



\section*{Acknowledgement}

I am very grateful to an anonymous referee for pointing out some
missing details in the original version of the proof of Lemma \ref{lemma3},
which led to the significant improvement of the paper.



\begin{thebibliography}{999}

\bibitem{avramidi91}
I. G. Avramidi,  
{\it A covariant technique for the calculation of the one-loop effective action}, 
Nucl. Phys., {\bf B355} (1991) 712--754  

\bibitem{avramidi93} 
I. G. Avramidi,
{\it A new algebraic approach for calculating the heat kernel in gauge
theories}, Phys. Lett., {\bf B305} (1993) 27--34


\bibitem{avramidi95}
I. G. Avramidi,
{\it Covariant algebraic method for calculation of the low-energy 
heat kernel}, J. Math. Phys., {\bf 36} (1995) 5055-5070; 
Erratum: J. Math. Phys., {\bf 39} (1998) 1720




\bibitem{avramidi09}
I. G. Avramidi, {\it Heat kernel on homogeneous bundles over symmetric spaces}, 
Commun. Math. Phys., {\bf 288} (2009) 963-1006


\bibitem{avramidi15}
I. G. Avramidi,
{\it Heat Kernel Method and Its Applications}, Birk\-h\"auser, Basel (2015)

\bibitem{avramidi17}
I. G. Avramidi, 
{\it Quantum heat traces}, J. Geom. Phys., {\bf 112} (2017) 271--288

\bibitem{avramidi20a}
I. G. Avramidi, 
{\it Relative spectral invariants of elliptic operators on manifolds}, 
J. Geom. Phys., {\bf 150} (2020) 103599.

\bibitem{avramidi20b}
I. G. Avramidi, {\it 
Bogolyubov invariant via relative spectral invariants on manifolds},
J. Math. Phys., {\bf 61}  (2020) 032303

\bibitem{avramidi16}
I. G. Avramidi and B. J. Buckman, 
{\it Heat determinant on manifolds}, J. Geom. Phys., {\bf 104} (2016) 
64--88


\bibitem{berger03} 
M. Berger, 
\textit{A Panoramic View of Riemannian Geometry},
Springer, Berlin (1992)


\bibitem{birrel80}
N. D. Birrell and P. C. W. Davies, 
{\it Quantum Fields in Curved Space},
Cambridge: 
Cambridge University Press, 1980

\bibitem{bonfiglioli12}
A. Bonfiglioli and R. Fulci,
{\it Topics in Noncommutative Algebra: The Theorem of Campbell, Baker, Hausdorff and Dynkin},
Springer, Berlin (2012)




\bibitem{dewitt2003}
B. S. Dewitt, The Global Approach to Quantum Field Theory, Volumes 1 and 2,
Oxford: Oxford University Press, 2003.

\bibitem{gilkey95}
P. B. Gilkey, 
{\it Invariance Theory, the Heat Equation and the Atiyah-Singer Index Theorem},
CRC, Boca Raton (1995)

\bibitem{gordon92}
C. Gordon, D. L. Webb and S. Wolpert,
{\it One cannot hear the shape of a drum},
Bull. Amer. Math. Soc. (N.S.) {\bf 27} (1992) 134-138

\bibitem{grib94}
A. A. Grib, S. G. Mamaev and V. M. Mostepanenko,
{\it Vacuum Quantum Effects in Strong Fields},
St. Petersburg: Friedmann Laboratory, 1994.





\bibitem{parker09}
L. Parker and D. Toms, {\it Quantum Field Theory in Curved Spacetime: Quantized Fields
and Gravity}, 
Cambridge: Cambridge Monographs on Mathematical Physics, 2009



\bibitem{prudnikov83} A. P. Prudnikov, Yu. A. Brychkov and O. I. 
Marychev,
{\it Integrals and Series}, vol. I,  CRC, Boca Raton (1998)

\bibitem{schueth99}
D. Schueth, {\it Continuous families of isospectral metrics on 
simply connected manifolds}, Ann. Math., {\bf 149} (1999) 287–308

\bibitem{wald94}
R. M. Wald, {\it Quantum Field Theory In Curved Space-Time And Black Hole Thermodynamics}, 
Chicago: University of Chicago Press, 1994



\end{thebibliography}
\end{document}